\documentclass[twocolumn,showpacs,amsmath,amssymb]{revtex4}

\usepackage{amsmath,amssymb,latexsym,graphics,epsfig}
\usepackage{pstricks}

\newpsobject{showgrid}{psgrid}{subgriddiv=1}

\newcommand{\td}{\mathrm{d}}
\newcommand{\zd}{\mbox{\boldmath $z$}}
\newcommand{\zds}{\mbox{\scriptsize\boldmath $z$}}
\newcommand{\ad}{\mbox{\boldmath $a$}}

\begin{document}

\title{Maximal height statistics for $1/f^{\alpha}$ signals}

\author{G. Gy\"{o}rgyi}
\email{gyorgyi@glu.elte.hu}
\author{N. R. Moloney}
\email{moloney@general.elte.hu}
\author{K. Ozog\'{a}ny}
\email{ozogany@general.elte.hu}
\author{Z. R\'{a}cz}
\email{racz@general.elte.hu}

\affiliation{Institute for Theoretical Physics - HAS,
  E\"{o}tv\"{o}s University, P\'{a}zm\'{a}ny
  s\'{e}t\'{a}ny 1/a, 1117 Budapest, Hungary}
\date{\today}

\begin{abstract}
Numerical and analytical results are presented for the maximal
relative height distribution of stationary periodic Gaussian signals
(one dimensional interfaces) displaying a $1/f^{\alpha}$ power
spectrum. For $0\le\alpha<1$ (regime of decaying correlations), we
observe that the mathematically established limiting distribution
(Fisher-Tippett-Gumbel distribution) is approached extremely slowly as
the sample size increases. The convergence is rapid for $\alpha>1$
(regime of strong correlations) and a highly accurate picture gallery
of distribution functions can be constructed numerically.  Analytical
results can be obtained in the limit $\alpha \to \infty$ and, for
large $\alpha$, by perturbation expansion. Furthermore, using path
integral techniques we derive a trace formula for the distribution
function, valid for $\alpha=2n$ even integer. From the latter we
extract the small argument asymptote of the distribution function
whose analytic continuation to arbitrary $\alpha >1$ is found to be in
agreement with simulations. Comparison of the extreme and roughness
statistics of the interfaces reveals similarities in both the small
and large argument asymptotes of the distribution functions.
\end{abstract}

\maketitle

\section{Introduction}
Whereas the extreme value statistics (EVS) of independent and
identically distributed (i.i.d.) random variables has been thoroughly
understood for a long
time~\cite{FisherTippett:1928,Gnedenko:1943,Gumbel:1958}, our
knowledge about the EVS of correlated variables is less general.  Many
natural processes, like flood-water levels, meteorological parameters,
and earthquake
magnitudes~\cite{KatzParlangeNaveau:2002,StorchZwiers:2002,GutenbergRichter:1944},
are, however, characterized by large variations, a phenomenon
connected to long term correlations. Since extremal occurrences in
physical quantities may be of great significance, it is essential to
develop an understanding of EVS in the presence of correlations.  The
last few years have seen increased activity in this direction, with
several particular cases worked out in detail. For example, extremal
height fluctuations in 1+1 dimensional Edwards--Wilkinson surfaces
have been investigated
recently~\cite{MajumdarComtet:2004,MajumdarComtet:2005}, and a
nontrivial distribution function, the Airy distribution, was found
analytically for the stationary surface.  Equivalently, considering
the latter as a time signal, this result relates to maximal
displacements in Brownian random walks.  Other studies of surface
fluctuations also demonstrate the effect of correlations on EVS, and
several examples show that nontrivial EVS may emerge even in the
simplest surface evolution
models~\cite{AntalETAL:2001,GyorgyiETAL:2003,
Lee:2005,BolechRosso:2004,Bertin:2005,GucluKorniss:2004}.
Remarkable connections have also been found between EVS and
propagating front solutions, exploited in such problems as random
fragmentation~\cite{KrapivskyMajumdar:2000}, or random binary-tree
searches~\cite{MajumdarKrapivsky:2002}. Correlations have also been
shown to play an important role in effecting extreme events in weather
records~\cite{BundeETAL:2005,KiralyBartosJanosi:2006}.
To summarize, problems related to extremes regularly arise, and it is
a fundamental question whether they obey a limit distribution
characterizing i.i.d.\ variables, or some special, nontrivial,
statistics emerges.

In order to develop an intuition about the effect of correlations, we
shall consider here the EVS of periodic signals displaying Gaussian
fluctuations with $1/f^\alpha$ power spectra.  While we shall use the
terminology of time signals,
one-dimensional stationary interfaces may equally be imagined,
with the same spatial spectrum,
and periodic boundary conditions. Systems with $1/f^\alpha$ type
fluctuations are abundant in nature, with examples ranging from voltage
fluctuations in resistors~\cite{YakimovHooge:2000}, through
temperature fluctuations in the oceans~\cite{MonettiHavlinBunde:2002}, to climatological temperature records~\cite{BlenderFraedrichHunt:2006},
to the number of stocks
traded daily~\cite{LilloMantegna:2000}.  In addition, most of these
fluctuations appear to be Gaussian, thus our results may have
relevance in answering questions about the probability of extreme
events therein.

The $1/f^\alpha$ signals we consider are rather simple in the sense
that they decompose into independent modes in Fourier space.  The
modes are not identically distributed, however, giving rise to
temporal correlations, which are by now well understood (see
Sec.~\ref{Gauss-1falpha}).  Correlations are tuned by $\alpha$,
yielding signals with no correlations ($\alpha =0$), decaying
($0<\alpha <1$), and diverging correlations ($1\le \alpha <\infty$).
Thus $1/f^\alpha$ processes are also well suited for studying the effect of
a wide range of correlations on extreme events in signals.

The central quantity we investigate is the maximum relative height
(MRH), first studied in~\cite{RaychaudhuriETAL:2001}.  This is
the highest peak of a signal over a given time interval $T$,
measured from the average level. Specifically, for each realization of
the signal, $h(t)$, the MRH is
\begin{equation}
  h_m := \max_t h(t) - \bar{h}(t) \, .
\label{E:MRH}
\end{equation}
where $\max_t h(t)$ is the peak of the signal and $\bar{h}(t)$ is its
time average.  The MRH, $h_m$, varies from realization to realization,
and is therefore a random variable whose probability density function
(PDF), denoted by $P(h_m)$, we would like to determine.  The physical
significance of $h_m$ is obvious. For instance, in a corroding surface
it gives the maximal depth of damage or, in general, it is the maximal
peak of a surface.  To name another example, when natural water level
fluctuations are considered, it is related to the necessary dam height.

Since the Fourier components of the signal are independent variables,
it is relatively easy to generate $h_m$-s numerically and thereby
obtain sufficient statistics for sampling $P(h_m)$ (see
Sec.~\ref{technicalities}). Scanning through $0\le \alpha <\infty$
reveals that $\alpha_c=1$ separates two regions with distinct
behaviors in both the limiting functions and the convergence to them
as the signal length ($0\le t\le T$) tends to infinity. At $\alpha=0$,
the signal is made up of i.i.d.\ variables and the EVS is governed by
the FTG distribution, which is one of the three possible limit
distributions for i.i.d.\ variables in the traditional
categorization~\cite{Galambos:1978,DeHaanFerreira:2006}.  In fact,
this property extends to the whole $0\le \alpha < 1$
interval~\cite{Berman:1964}, where the correlations decay in a
power-law fashion. Our results indicate that, at least in the $0\le
\alpha < 0.5$ region, not only the limit distribution but the
convergence to it follows closely the logarithmically slow convergence
which characterizes $\alpha=0$ (Sec.~\ref{S:decaying-corr}). We find
that the convergence further slows down in the $0\le \alpha < 0.5$
region and it remains an open question whether it is slower or not
than logarithmic.

For $\alpha >1$, the signal becomes rough, that is, the correlations
diverge with signal length, and we find that the qualitative features
of the EVS in this range are the same as in the $\alpha =2$
case (Sec.~\ref{S:exact-results}), exactly solved by Comtet and
Majumdar~\cite{MajumdarComtet:2004,MajumdarComtet:2005}. Namely, the
divergent scale of the extreme values $\langle h_m \rangle \sim
T^\beta$, where $\beta=(\alpha-1)/2$, is proportional to the scale of
the fluctuations in the signal (square root of the roughness in
interface language) and, furthermore, the large- and small-argument
asymptotes of the limiting distribution functions are of similar
type. In order to demonstrate these similarities, we study the
generalized, higher order, random acceleration problem ($\alpha =$
even integer) in Sec.~\ref{sec:alpha-2n}, and calculate the
propagator of this process. Using this result, we develop a
generalization of the trace formula (Sec.~\ref{sec:mrh-pi}) which was
instrumental in solving the $\alpha =2$ problem. It turns out that the
trace formula can be written in a scaling form, which yields the scale
of the MRH values (Sec.~\ref{alpha.gt.1}) as well as, under a rather
mild and natural assumption, the small-argument asymptote of the MRH
distribution (Sec.~\ref{sec:small-x}).  Our numerical evaluations of
the distributions are all in excellent agreement with the analytical
results.

Analytical results can also be obtained in the $\alpha\to \infty$
limit (Sec.~\ref{S:exact-results}), where the lowest frequency mode
determines the shape of the signal. We find that the MRH distribution
has the functional form $\sim x\exp{(-x^2)}$. Corrections to the
$\alpha\to\infty$ limit may be obtained by keeping the lowest
frequency modes. With only three modes, a satisfactory description of
the whole $\alpha \ge 6$ region can be obtained
(Sec.~\ref{sec:large-alpha}).  Since both the $\alpha =2$ and $\alpha
=\infty$ results suggest that the large-argument tail of the
distribution takes the form $\sim x^\gamma\exp{(-x^2)}$, we checked
this property for other $\alpha$-s as well, and found it to be an
excellent description for all $\alpha >1$.

The common scaling properties of the maximal height and the root mean
square height for $\alpha >1$ lead us to compare the MRH distributions
to the roughness distributions of $1/f^\alpha$
interfaces~\cite{AntalETAL:2002}. We find in Sec.~\ref{sec:comp-rough}
that, in addition to the general shape of the PDF-s, both the small
and large argument asymptotes of these functions have analogous
functional forms provided the replacement $h_m\to (roughness)^{1/2}$
is made. Similar conclusions can also be reached when $P(h_m)$ is
compared with the distribution of maximal intensities~\cite{GyorgyiETAL:2003}.

Concluding remarks are collected in Sec.~\ref{summary} while details
of the calculations of the generalized random acceleration process and
of the large $\alpha$ expansion are given in Appendix
\ref{app:gf-background} and \ref{app:large-alpha}, respectively.

\section{Gaussian periodic $1/f^{\alpha}$ signals}
\label{Gauss-1falpha}

We consider Gaussian periodic signals $h(t)=h(t+T)$ of length $T$.
The probability density functional of $h(t)$ is given by
\begin{equation}
  \mathcal{P}[h(t)] \sim e^{-S[h(t)]},
  \label{E:action}
\end{equation}
where the effective action $S$ can be formally defined in real space
but, in practice, is defined through its Fourier representation
\begin{equation}
  S[c_n;\alpha] = 2\lambda T^{1-\alpha} \sum_{n=1}^{N/2} n^{\alpha} |c_n|^2\, .
\label{E:fourier_action}
\end{equation}
Here $\lambda$ is a stiffness parameter which is set to
$(2\pi)^\alpha /2$ hereafter (for the details and notation we follow
\cite{AntalETAL:2002}), and the $c_n$-s are the Fourier
coefficients of $h(t)$
\begin{equation}
  h(t) =\sum_{\small {n=-N/2+1}}^{N/2} c_ne^{2\pi i n t/T} \, ,
\label{E:fourier_transform}
\end{equation}
where $c_n^*=c_{-n}$ and their phases (for $n\not= N/2$)
are independent random variables uniformly
distributed in the interval $[0,2\pi]$, while $c_{N/2}$ is real.
Since $c_0$ does not appear in the action \eqref{E:fourier_action}
we can set the average of the signal to zero, i.e. $c_0=0$.
Note that the cutoff introduced by $N$ means that the
timescale is not resolved below
\begin{equation}
\tau=T/N
\end{equation}
and thus a measurement of $h(t)$ yields
effectively $N$ data points.

As one can see from Eqs.~\eqref{E:action} and
\eqref{E:fourier_action}, the amplitudes of the Fourier modes are
independent, Gaussian distributed variables -- but they are not
identically distributed. Indeed, the fluctuations increase with
decreasing wavenumber, with power spectrum
\begin{equation}
  \langle |c_n|^2 \rangle \propto \frac{1}{n^{\alpha}},
\end{equation}
as befitting a $1/f^{\alpha}$ signal.

By scanning through $\alpha$, systems of wide interest may be
generated. For example, $\alpha = 0,1,2,4$ correspond respectively to
white-noise, $1/f$-noise~\cite{Weissman:1988}, an Edwards-Wilkinson
interface~\cite{EdwardsWilkinson:1982} or Brownian curve, and a
Mullins-Herring interface~\cite{Mullins:1957,Villain:1991} or random
acceleration process~\cite{Burkhardt:1993}.

\begin{figure}
\includegraphics{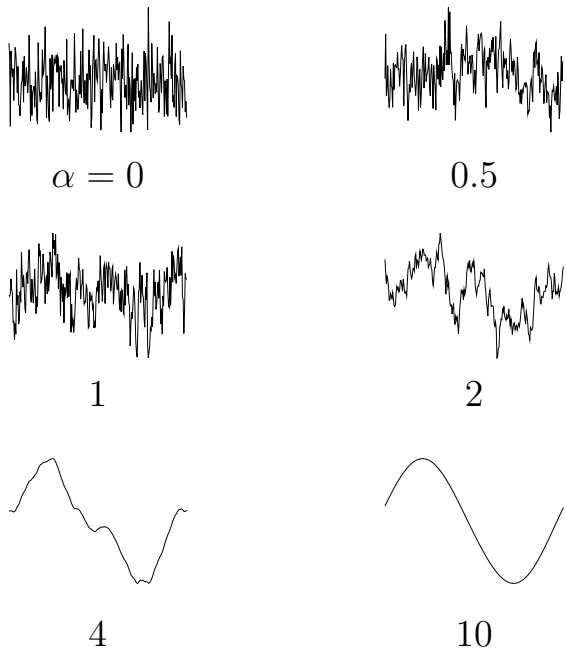}
\caption{Typical profiles of various $1/f^\alpha$ signals of length
  $N=T/\tau=8192$. Note that, contrary to the visual illusion, the
  $0\le\alpha<1$ surfaces are flat, while the $1\le\alpha<\infty$
  signals are rough. In the former case, the amplitude of the signals
  is size independent, while in the latter case the amplitude diverges
  with system size. For ease of comparison, we have
  rescaled the signals to be approximately equal in height.
  \label{F:alpha_scan}}
\end{figure}

An important feature of $1/f^\alpha$ signals is that correlations may
be tuned by the parameter $\alpha$.  Indeed, as one can see in
Fig.\ref{F:alpha_scan}, an $\alpha$-scan leads us from the absence of
correlations ($\alpha=0$, white noise) to the limit of a
deterministic signal ($\alpha=\infty$).  In between
$\alpha=0$ and $\alpha\to\infty$, $\alpha_c=1$ separates decaying
$(0\le\alpha<1)$ and strongly-correlated $(1\le\alpha<\infty)$
signals.
As we shall see, the extreme statistics is different in these two
regions, thus it may be worth spelling out the distinctions between
decaying and strong correlations. We therefore briefly describe some
known results regarding the correlations in $1/f^\alpha$ signals that
will be relevant to the understanding of the rest of the paper.

A global (integral) characteristic of correlations is given by the
mean-square fluctuations of the signal, called roughness or width in surface
terminology~\cite{Krug:1997,BarabasiStanley:1995}
\begin{equation}
  w_2=\overline{[\, h(t)-{\overline{h}}\,]^2} =
  2 \sum_{n=1}^{N/2} |c_n|^2 \, ,
  \label{w2-eq}
\end{equation}
where the overbar indicates an average over $t$, and the second
equality shows that $w_2$ is the integrated power-spectrum of the
system. This quantity has been
much investigated~\cite{FoltinETAL:1994,AntalETAL:2002}
and its probability distribution will be compared to
the extreme statistics of the surface in Sec.~\ref{sec:comp-rough}.
For the present purpose it is sufficient to recall that the ensemble
average over surfaces, $\langle w_2\rangle$, yields the following
asymptote for large system sizes ($T\to \infty$)
 \begin{equation}
 \langle w_2 \rangle\sim
\left
\{\begin{array}{ll}
\, T^{\alpha-1} & \mbox{ for } 1 < \alpha \le \infty\\
\, \ln{T/\tau} & \mbox{ for }  \,\, \alpha =1 \\
\, \tau^{\alpha-1} & \mbox{ for } 0\le \alpha <1\,\, .
\end{array}
\right.
\label{alt}
\end{equation}
Thus the fluctuations diverge with system size for $1\le \alpha
<\infty$ in contrast to the finite fluctuations in the $0\le \alpha
<1$ regime.  Since diverging fluctuations are the sign of strong
correlations, this gives a reason for separating the $0\le \alpha <1$
and $1\le \alpha <\infty$ regions and attaching the name of decaying
and strong correlations to each, respectively.

A more detailed characterization of the $\alpha$-dependence of the
correlation can be obtained by examining the correlation function
$C_\alpha(t,T)=\langle h(t^\prime)h(t^\prime +t)\rangle$ itself.  A
simple calculation shows that the limit $T\to \infty$ and $t/T\to
finite$ yields the following scaling form
\begin{equation}
  C_\alpha(t,T)=T^{\alpha -1}F_\alpha(t/T) \, ,
\end{equation}
and that the nature of the correlations follows from the properties of
scaling function $F_\alpha$.

For $1<\alpha<\infty$, the scaling function is of order ${\cal O}(1)$
and $F_\alpha(t/T\to 0)$ is finite. As a consequence,
\begin{equation}
  C_{\alpha>1}(t,T)\sim T^{\alpha-1} \, ,
\end{equation}
so that the correlations diverge in the $T\to \infty$ limit.  The
divergence is also present for $\alpha=1$ but it is only logarithmic,
$C_1(t,T)\sim \ln{(T/\tau)}$.  Systems with $1\le\alpha<\infty$ can
therefore be regarded as {\em strongly correlated}.

For $0<\alpha<1$, the correlations are ${\cal O}(1)$ since the scaling
function behaves as $F_\alpha(u)\sim u^{1-\alpha}$ for $u\ll 1$ and,
consequently, one has a power law decay of correlations,
independent of system size
\begin{equation}
  C_{\alpha<1}(t,T)\sim 1/t^{1-\alpha} \, .
\end{equation}
In the bulk $(u\sim 1/2)$, the correlations quickly approach zero,
$C_\alpha\sim 1/T^{1-\alpha}$ in the $T\to\infty$ limit. The
correlations disappear entirely for $\alpha=0$ since, in this case,
$h(t)$ are i.i.d.\ variables.  Systems with $0\le\alpha<1$ have {\em
decaying correlations} hence the name used for their identification.

Thus we see how the regions $0\le\alpha<1$ and $1\le\alpha<\infty$ are
distinguished. Furthermore, we also have a characterization of
correlations taken into account when we study the EVS of periodic
Gaussian $1/f^\alpha$ signals.

\section{Extreme statistics: technicalities}
\label{technicalities}

The quantity of interest is the distribution function $P(h_m)$ of the
maximum height $h_m$ of the signal measured from the average,
as defined in Eq.~\eqref{E:MRH}. In order to construct the
histogram for the frequency distribution of $h_m$, we generate a large
number $(\approx 10^6-10^7)$ of signals, as prescribed by the action
$S[c_k;\alpha]$ in Eq.~\eqref{E:fourier_action}. Each signal is
Fourier transformed and the real-space signal, which has zero average
($c_0=0$), is used to determine the value of $h_m$.  Finally, the
$h_m$-s are binned to build the histogram for the MRH distribution.

Since $h_m$ is selected as the largest from $N=T/\tau$ numbers,
$P(h_m)$ obtained by the above recipe depends on $N$.  The goal of EVS
is to find the limiting distribution which emerges for $N\to \infty$
\begin{equation}
 P(z)=\lim_{_{N\to\infty}}a_{_N}P_{_N}(h_m=a_{_N}z+b_{_N}) \, .
\end{equation}
Here $a_N$ and $b_N$ are introduced to take care of the possible
singularities in $\langle h_m\rangle_N$ and in $\sigma_{_N}^2=\langle
(h_m - \langle h_m \rangle)^2\rangle$ (one expects e.g. that $\langle
h_m\rangle_{N\to\infty}\to \infty$ for distributions with no finite
upper endpoint).

For any finite $N$, the parameters $a_{_N}$ and $b_{_N}$ can be
related to $\langle h_m\rangle_{_N}$ and $\sigma_{_N}$ and, in
practice, one builds a scaled distribution function where $a_{_N}$ and
$b_{_N}$ do not play any role.  In the following, we shall employ two
distinct scaling procedures.  If the large $N$ behaviors of $\langle
h_m\rangle_{_N}$ and $\sigma_{_N}$ coincide (e.g. $ \langle
h_m\rangle_{_N}\sim\sigma_{_N}\sim N^\theta$) then we use {\em scaling
by the average} by introducing the variable
\begin{equation}
x=h_m/\langle h_m\rangle_{_N}
\end{equation}
which ensures that $\langle x\rangle=1$ and makes the corresponding
scaling function
\begin{equation}
\Phi(x)=\lim_{_{N\to\infty}} \Phi_{_N}(x)=\lim_{_{N\to\infty}} \langle
h_m\rangle_{_N}P_{_N}(\langle h_m\rangle_{_N}x)
 \,
\end{equation}
devoid of any fitting parameters.

If $\langle h_m\rangle_{_N}$ and $\sigma_{_N}$ scale differently in
the large $N$ limit then the above procedure leads either to a delta function
or to an ever widening distribution. One can deal with this problem by
measuring $h_m$ from $\langle h_m\rangle$ in units of the standard
deviation i.e. by introducing the scaling variable
\begin{equation}
  y = \frac{h_m - \langle h_m \rangle}{\sqrt{\langle h^2_m \rangle -
  \langle h_m \rangle^2}}=(h_m - \langle h_m \rangle)/\sigma_{_N} \, .
\end{equation}
Using $y$ will be called {\em $\sigma$-scaling} and the corresponding
scaling function will be denoted by $\tilde \Phi (y)$.  Provided the
limit $N\to \infty$ exists,
\begin{equation}
 \tilde\Phi(y)=\lim_{_{N\to\infty}}\tilde\Phi_{_{N}}(y)=\lim_{_{N\to\infty}}
 \sigma_{_N} P_{_N}(\langle h_m\rangle_{_N}+\sigma_{_N}y)
\end{equation}
is again a function without any fitting parameters.

\section{EVS in the regime of decaying correlations ($0\le \alpha <1$)}
\label{S:decaying-corr}

In the white-noise limit $\alpha \to 0$, each point on the signal
constitutes a random i.i.d.\ variable with Gaussian
distribution.  Under these conditions, the MRH limiting distribution
falls under the domain of attraction of the Fisher-Tippett-Gumbel
distribution~\cite{FisherTippett:1928,Gnedenko:1943}. In fact, in the
range $0 \le \alpha < 1$, it has been shown that the decaying
correlations are too weak to change the FTG
limit~\cite{Berman:1964}. Therefore, in the regime of decaying
correlations, the MRH statistics of $1/f^{\alpha}$ signals may be said
to be universal.
\begin{figure}[htb]
\includegraphics{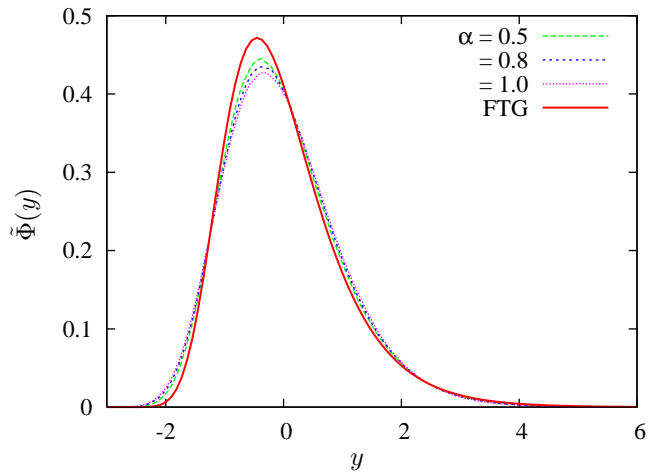}
\caption{Numerically constructed MRH distributions for $\alpha =
  0.5,0.8$ and $1.0$ for system size $N=T/\tau =16384$. The FTG
  distribution is shown by the thick black line. Each distribution is
  rescaled to zero mean and unit standard deviation. It should be
  mentioned that the $\alpha<0.5$ curves are not displayed since they
  are indistinguishable from the $\alpha=0.5$ case.
  \label{F:ftg_alphas01}}
\end{figure}

However, in the case of i.i.d. random variables drawn from a Gaussian
parent distribution (i.e. for $\alpha = 0$), it has also been
established that the convergence in $N$ towards the limiting FTG
distribution is logarithmically
slow~\cite{FisherTippett:1928,DeHaanResnick:1996,GomesDeHaan:1999}.
Therefore, in
practice, the MRH distribution may appear different from FTG.  An even
worse rate of convergence may be expected with increasing $\alpha$,
since, heuristically, increasing correlations decrease
the effective number of degrees of freedom. In
Fig.~\ref{F:ftg_alphas01} we illustrate this trend by comparing
numerical MRH distributions for a range of $\alpha$ but fixed $N$ with
the FTG limiting distribution. For $\alpha < 0.5$ the numerical
distributions are practically indistinguishable from the case $\alpha
= 0.5$. This figure serves as a warning when comparing real-world data
with known extreme value distributions.

We note here that Eichner et al.~\cite{EichnerETAL:2006} have recently
investigated EVS for $\alpha =0.6$. Although they do not spell it out
explicitly, their Figs. 2 and 3 do demonstrate that the convergence
at $\alpha =0.6$ (Fig.\ 3) is slower than at $\alpha =0$ (Fig.\ 2), in
agreement with our findings described above.

\begin{figure}[htb]
\includegraphics{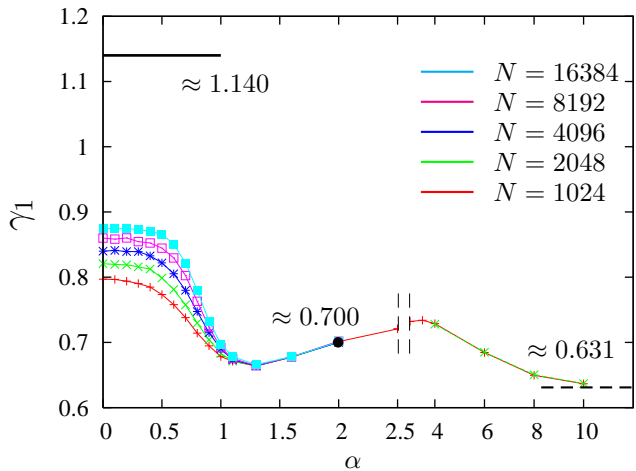}
\caption{Strong finite-size effects at low values of $\alpha$ as seen
in the scaled skewness, $\gamma_1=\kappa_3/\kappa_2^{3/2}$ of the MRH
distribution. Note the exponential
increasing system sizes $N=T/\tau$ used for comparisons. Note also that
the scale of the horizontal axis changes at $\alpha =2.5$.
The limiting value of $\gamma_1$ for $\alpha< 1$ is
$\gamma_1\approx 1.140$.  Other exactly known values are
$\gamma_1(\alpha=2)\approx 0.700$ and
$\gamma_1(\alpha\to\infty)\approx 0.631$.}
\label{F:skewness}
\end{figure}

In order to shed more light on convergence rates towards limiting
distributions, we have measured the skewness $\gamma_1 =
\kappa_3/\kappa_2^{3/2}$ where $\kappa_n$ is the $n$-th cumulant of
the MRH distribution function. The results for a range of $\alpha$ and
$N$ are displayed in Fig.~\ref{F:skewness}.  From this plot one
can discern a number of remarkable features. First, in the range $0
\le \alpha < 1$, we note that the measured skewnesses are far from the
skewness of the FTG distribution (approximately $1.140\ldots$), even
for the largest system size available. Second, for $\alpha = 0$, we
know theoretically that the convergence rate is logarithmically slow,
but, somewhat surprisingly, this convergence rate appears to be shared
for all $\alpha \leq 0.5$, after which convergence slows down
markedly. Thus, the
universality in the ultimate limiting distribution for $0 \le \alpha <
1$ may not carry over to a universality in the finite-size
corrections. Note that if we did not know the limit but would try to
determine it from finite-N skewnesses, then for
$\alpha \lesssim 1$ we would be wrong to conclude that
the asymptotic value had nearly been reached.

The case of strong correlations is discussed in the following
sections.  Here, we just observe that the skewnesses for $\alpha > 1$
rapidly collapse for different $N$, and that they are virtually
indistinguishable from each other for $\alpha \gtrsim 1.5$.  In this
case we may be quite sure that the skewnesses have practically reached
their limiting values, since they match their corresponding
theoretical values for $\alpha = 2$ and $\infty$ with high
accuracy. As we shall argue in Section~\ref{alpha.gt.1}, in contrast
to the very slow convergence for $0\le\alpha<1$,
convergence rate improves as it becomes a power law for $\alpha > 1$.

\section{Strong-correlation regime:
Exact results for $\alpha =2$ and $\alpha =\infty$.}
\label{S:exact-results}

The $1\le \alpha <\infty$ region is characterized by diverging
mean-square fluctuations (see Eq.~\eqref{alt}).  Since $\langle
h_m\rangle\ge\sqrt{\langle w_2\rangle}$, this is also a range where
the characteristic scale of $\langle h_m\rangle$ diverges with the
size of the system at least as $\langle h_m\rangle \sim T^{(\alpha -
1)/2}$.  An important exact result in the strongly-correlated regime
is related to the Brownian random walk ($\alpha=2$). Majumdar
and Comtet~\cite{MajumdarComtet:2004,MajumdarComtet:2005} have shown
that $\langle h_m\rangle \sim \sqrt{T}$ and, furthermore, they
calculated the MRH distribution using path-integral techniques as well
as by making a mapping to the problem of the area distribution under a
Brownian excursion~\cite{Takacs:1991,Takacs:1995}.  The resulting
distribution is known as the Airy distribution.  Under {\em scaling by
the average} ($x=h_m/\langle h_m \rangle$), the Airy distribution can
be written as follows (note that slightly different scaling has been
used in~\cite{MajumdarComtet:2004,MajumdarComtet:2005})
\begin{equation}
\Phi(x)=\frac{8\sqrt{3}}{\sqrt{\pi}x^{10/3}}
\sum_{n=1}^{\infty}e^{-v_n/x^2}v_n^{2/3} U(-5/6,4/3,v_n/x^2)
 \, .
 \label{Majumdar-full}
\end{equation}
Here $U(a,b,z)$ is the confluent hypergeometric function and
$v_n=(2/\pi)\cdot (2\alpha_n/3)^3$ is related to the $n$-th zero
$-\alpha_n$ of the Airy function.

The small and large $x$ asymptotes of $\Phi(x)$ have also been
calculated~\cite{MajumdarComtet:2004,MajumdarComtet:2005} with the
results
\begin{equation}
\Phi(x \to 0)\sim \frac{8\sqrt{3}v_1^{3/2}}{\sqrt{\pi} x^5} e^{-v_1/x^2}
 \,
 \label{small-x-asymp}
\end{equation}
and
\begin{equation}
\ln \Phi(x \to \infty)\sim {-3\pi x^2/4}
 \, .
 \label{large-x-asymp}
\end{equation}
It is noteworthy that the above asymptotes are quite close in
functional form to those obtained for the width distribution of the
Edwards-Wilkinson model~\cite{FoltinETAL:1994}.

\begin{figure}[htb]
\includegraphics{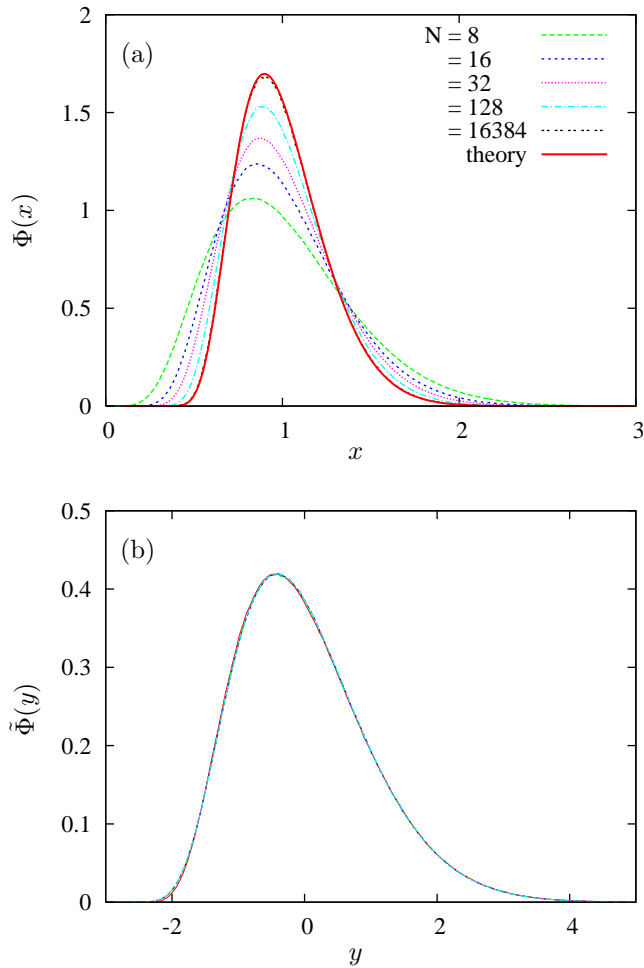}
\caption{Convergence of the MRH distribution to its exactly known
$N\to \infty$ limit~\cite{MajumdarComtet:2004}. Results for same
system sizes are displayed in both panels using scaling by the average
($x=h_m/\langle h_m\rangle $) and $\sigma$-scaling [$y=(h_m - \langle
h_m\rangle)/\sigma$] in the upper and lower panels, respectively.
Very small system sizes were also displayed in order to demonstrate
the remarkably fast convergence by $\sigma$-scaling.
\label{F:mu_sigma_scaling}}
\end{figure}

The plot of $\Phi(x)$ is shown on Fig.\ \ref{F:mu_sigma_scaling}a where
a rather fast convergence to the limiting function can be seen (the
convergence rate is $T^{-1/2}$ as calculated
in~\cite{SchehrMajumdar:2006}).  It is remarkable that the convergence
is even faster if {\em $\sigma$-scaling} is used
(Fig.\ \ref{F:mu_sigma_scaling}b). The reason for this is the
finite-size scaling of higher cumulants of the MRH distribution
function. At this point we present this just as a numerical
observation. A detailed study of the finite-size scaling of MRH will
be published separately~\cite{GyorgyiETAL:2006}.

The review of the properties of the MRH distribution for $\alpha =2$
presented above gives a guidance for the discussion of EVS in the
strong correlation regime. As we shall see below, the basic properties
of EVS ($\langle h_m\rangle\sim \sqrt{\langle w_2\rangle }$, the
general shape of $\Phi(x)$, the structure of the small- and large-$x$
asymptotics of $\Phi(x)$, the fast convergence to the limiting
function) are similar in the whole $1<\alpha<\infty$ region.

The other analytically solvable case is the $\alpha\to\infty$ limit.
Indeed, here only the $n=1$ mode survives, and the resulting signal
$h(t)=|c_1|\sin{(2\pi t/T+\varphi)}$.  Consequently, $h_m=|c_1|$
and the distribution of $h_m$ is just the distribution of $|c_1|$
given by $P(|c_1|)\sim |c_1|\exp{[-(2\pi)^\alpha T^{1-\alpha}
|c_1|^2]}$. Using the {\em average scaling}, the scaling function
$\Phi_{_\infty} (x)=\langle h_m \rangle P(\langle h_m \rangle x)$
becomes
\begin{equation}
 \Phi_{_\infty}(x)= \frac{\pi}{2} x\theta(x)e^{-\pi x^2/4} \, .
  \label{a-inf-pdf}
\end{equation}
were $\theta(x)$ is Heaviside's step function.  Comparing the above
expression with the asymptotes \eqref{small-x-asymp} and
\eqref{large-x-asymp}, one can see that, in addition to the
disappearance of the small $x$ singularity, the large $x$ asymptote has
also changed by an extra $x$ factor.

\section{The propagator of the generalized random acceleration process:
$\alpha=2n$}
\label{sec:alpha-2n}

The derivation of subsequent analytical results on the scale of
$\langle h_m\rangle$,
and on the small $h_m$
asymptote of the MRH distribution is based on the observation that
$1/f^\alpha$ signals are actually paths of generalized random
acceleration processes, provided that $\alpha =2n$ is an even
integer. This allows a path-integral representation of the MRH
distribution function (Sec.~\ref{sec:mrh-pi}) from which rather general
conclusions can be drawn and, furthermore, as indicated by the
simulations, the results can be extended to any $1<\alpha <\infty$.

The construction of the MRH distribution function in the path integral
approach involves the calculation of a normalization factor which, in
turn, requires the knowledge of the propagator (also called two-time
Green-function, or transition probability) of the random acceleration
process.  Here we compute this propagator, i.e.  the probability density
of a position of the stochastic path at some time $t$ conditioned on the
initial point.

The equation of motion of the $1/f^{2n}$ trajectory reads
\begin{eqnarray}
  h^{[n]}(t) = \frac{\td^nh(t)}{\td t^n} = \xi(t),
  \label{eq:gen-rnd-acc}
\end{eqnarray}
where $\xi(t)$ is white noise with zero mean and correlation
$\left\langle \xi(t)\xi(t')\right\rangle=\delta(t-t')$. Note that
$h(t)$ corresponds to the $n-1$-st integral of the Brownian random walk
trajectory. Eq.\ \eqref{eq:gen-rnd-acc} can be
rewritten as a vector Langevin equation
\begin{eqnarray}
  \dot z_1 = \xi(t),\ \ \dot z_k = z_{k-1}, \ (k=2,3,\dots n), \ \
  z_n=h.
  \label{eq:z-x}
\end{eqnarray}
For $n=1$ we have the usual random walk, for $n=2$ the random
acceleration problem~\cite{Burkhardt:1993}, also extensively studied,
while for higher $n$-s one can speak about the generalized random
acceleration processes~\cite{MajumdarBray:2001,SchwarzMaimon:2001}.  We
are interested in the conditional probability that after time $t$ the
trajectory is at $\zd=(z_1,z_2,\ldots,z_n)$ provided it started from
$\zd^0$. In the following, we denote this propagator by
$G_n(\zd|\zd^0;t)$.  Its subscript indicates the dimension of the
vector arguments, and it obviously satisfies the recursion relation
\begin{eqnarray}
  G_n(\zd|\zd^0;t) = \int\td z_{n+1} \, G_{n+1}(\zd|\zd^0;t)
\label{eq:GF-int}
\end{eqnarray}
where the integration eliminates the dependence on $z_{n+1}^0$, too.
The propagator has Dirac delta initial condition,
$G_n(\zd|\zd^0;0)=\delta^{(n)}(\zd-\zd^0)$, and satisfies the
Fokker-Planck equation, obtained in a standard way from the Langevin
equation~\cite{VanKampen:1992},
\begin{eqnarray}
  \partial_t G_n  &=& - \hat H_n^0 G_n \label{eq:FP0-n}\\
  \hat H_n^0 &=& -\frac 12 \partial_1^2 +\sum_{k=1}^{n-1} z_k
  \partial_{k+1},
  \label{eq:H0}
\end{eqnarray}
where $\partial_t$ and $\partial_k$ are derivatives with respect to
$t$ and $z_k$, respectively. The superscript of $\hat H_n^0$
refers to the fact
that we consider here the time evolution \eqref{eq:gen-rnd-acc}
without further constraints.  The propagator has been calculated in
previous studies up to $n=
5$~\cite{ChaichianDemichev_1:2001,SchwarzMaimon:2001}.  We make an ansatz
that matches these functions and we show it to be valid for general
$n$
\begin{eqnarray}
    G_n(\zd|\zd^0;t) = \prod_{k=1}^n \, \mathcal{G}(\ad^k \!\!\cdot\! \zd -
    \ad^{0,k} \!\!\cdot\! \zd^{0};\sigma_k),
    \label{eq:gf-ansatz}
\end{eqnarray}
where $\mathcal{G}(z;\sigma)=\exp{(-z^2/2\sigma^2 )}/\sqrt{2\pi}\sigma$
is a Gaussian PDF with zero mean and
variance $\sigma^2$, $\zd^{0}$ is the initial condition vector, and
the vector $\ad^k$ only has nonzero components for $i=1,2,\dots, k$,
i.e., for $l>k$ we have $a_l^k=0$. In order to remove ambiguity we set
$a_k^k=1$. The $\ad^k, \ad^{0,k}, \sigma_k$ are time dependent
quantities to be determined.  The above formula amounts to the
recursion relation
\begin{eqnarray}
  G_n(\zd|\zd^{0};t) &=& G_{n-1}(\zd|\zd^{0};t) \nonumber \\
  && \times \, \mathcal{G}(\ad^n \!\!\cdot\! \zd - \ad^{0,n} \!\!\cdot\!
\zd^{0};\sigma_n).
\label{eq:gf-rec}
\end{eqnarray}
Substitution of this ansatz into \eqref{eq:FP0-n} leads to equations
for the unknown parameters. The solution of the equations as described
in Appendix \ref{app:gf-background} yields
\begin{subequations}
\begin{eqnarray}
  a_k^n \,\,&=& \frac{(-t)^{n-k} (n+k-2)!}{ 2^{n-1} (2n-3)!! (k-1)! (n-k)!}, \\
  a_k^{0,n} &=& (-1)^{n-k} a_k^n, \\
  \sigma_n\,\, &=& \frac{t^{(2n-1)/2}}{ 2^{n-1} \sqrt{2n-1} (2n-3)!!}\, .
\end{eqnarray}
\label{eq:gf-parms}
\end{subequations}

For illustration, we use the above expressions to calculate and
display explicitly the $n=4$ propagator.  Noting that the original
coordinate, velocity, acceleration, and its time derivative are given
by
\begin{eqnarray}
  h=z_4, \ \ v=z_3, \ \ a=z_2, \ \ \dot a=z_1,
\end{eqnarray}
respectively, the propagator can be written in the form
\begin{eqnarray}
  G_4(\zd|\zd^{0};t) = \frac{720\sqrt{105}}{\pi^2t^8} e^{-A/2},
\end{eqnarray}
where $-A/2$ is the sum of the exponents of the Gaussians in
\eqref{eq:gf-ansatz}, namely
\begin{equation}
A = \sum_{k=1}^4 A_k,
\end{equation}
with
\begin{subequations}
\begin{eqnarray}
 A_1 &=&\frac 1t (z_1-z_1^0)^2, \\
  A_2 &=& \frac{12}{t^3} \left[ z_2-z_2^0 - \frac t2 (z_1+z_1^0)\right]^2,  \\
  A_3 &=& \frac{720}{t^5} \left[ z_3-z_3^0 - \frac t2 (z_2+z_2^0) +
    \frac{t^2}{12} (z_1-z_1^0)\right]^2, \\
  A_4 &=& \frac{100800}{t^7} \left[z_4-z_4^0 -\frac t2 (z_3+z_3^0) +
    \frac{t^2}{10} (z_2-z_2^0) \right., \nonumber \\
  && \left. -\frac{t^3}{120} (z_1+z_1^0) \right]^2.
\end{eqnarray}
\end{subequations}
Up to the $k=3$ term this incorporates the propagators of the random
walk, $k=1$, random acceleration, $k\leq 2$, and random velocity of
acceleration, $k\leq 3$, and the above expressions are in agreement
with previous results~\cite{SchwarzMaimon:2001}.  Note that,
independently of $k$ we have $a_k^k=a^{0,k}_k=1$, and
\begin{eqnarray}a^k_{k-1}=-a^{0,k}_{k-1}=t/2,
  \label{eq:akk-1}\end{eqnarray}
but for $l\leq k-2$ the $a_l^k$-s will vary with both $l$ and $k$.

Later, for the construction of the formula for the MRH distribution,
we will need a special property of the propagator.  Namely, if we
consider the propagator of a periodic path of length $T$ and integrate
it over the common values of the velocity, acceleration, etc., at the
endpoints, we get the surprisingly simple result
\begin{eqnarray}
  \int \prod_{k=1}^{n-1}\td z_k^0 \, G_n(\zd^0|\zd^{0};T) =
  T^{-(n-1/2)}  (2\pi)^{-1/2}.
  \label{eq:int-per-gf}
\end{eqnarray}
Indeed, the periodic propagator does not depend on $z_n^0$, the
integration over $z_{n-1}^0$ cancels the normalizing constant of the
$n$-th Gaussian but brings in a factor of $1/T$. The integration over
$z_{n-2}^0$ does the same with the $n-1$-st Gaussian, and so on, until
finally we are left with the norm factor of the $n=1$ Gaussian,
$1/\sqrt{2\pi T}$, divided by $T^{n-1}$, as shown in
\eqref{eq:int-per-gf}.  The key to this remarkable cancellation of the
total numeric prefactor of the propagator is that \eqref{eq:akk-1}
holds uniformly for all $k$-s.

\section{Path integral formalism and the trace formula
for the MRH ($\alpha=2n$)}
\label{sec:mrh-pi}

For $\alpha=2$ Majumdar and
Comtet~\cite{MajumdarComtet:2004,MajumdarComtet:2005} introduced a
path integral representation of the MRH distribution.  The
technique allowed for the formulation of the PDF in terms of the
spectrum of a quantum mechanical, one-dimensional, Hamiltonian $\hat
H$ with a hard wall and elsewhere linear potential, through the trace
of $e^{-\hat H T}$, valid in the case of periodic boundary conditions.
The spectrum is known to consist of the Airy zeros, so the trace
formula resulted in the PDF called the Airy distribution.

In what follows we show that, in the case of periodic boundary
conditions, for a general $\alpha=2n$, $n=2,3,\dots$ an analogous
trace formula holds.  Remarkably, the formula turns out to be
essentially the same as in the $\alpha=2$ case, with the only
difference that now a generalized ``Hamiltonian'' $\hat H_n$ appears.
However, the $\hat H_n$, a differential operator in an $n$ dimensional
space, is no longer Hermitian.  Whereas we shall not solve the
spectral problem necessary for the calculation of the MRH
distribution, this formulation will allow us to (i) determine the
scale of the MRH as function of $T$, and (ii) give explicitly the
initial asymptote of the PDF, with the only undetermined parameter
being the ground state energy of the Hamiltonian $\hat H_n$.  What is
more, the results (i-ii) will lend themselves to a continuation to
real $\alpha$-s, so the use of the path integral technique extends
beyond its original region of validity, the generalized random
acceleration problem $\alpha=2n$.

We begin with the probability functional of a periodic path $h(t)$,
where $h$ is measured from the time average,
\begin{eqnarray}
  P[h(t)] &=& \mathcal{A} \exp\left( - \frac 12 \int_0^T\td t\,
    [h^{[n]}(t)]^2 \right)\nonumber \\&&\times \delta\left(
    \int_0^T\td t \, h(t)\right).
  \label{eq:mrh-pi}\end{eqnarray}
Following~\cite{MajumdarComtet:2004,MajumdarComtet:2005} we have
introduced a normalizing coefficient $\mathcal{A}$, ensuring
\begin{eqnarray}
  \int_{\text{PBC}} \hspace{-10pt}\mathcal D_n h(t) P[h(t)] =1,
\end{eqnarray}
where PBC indicates that periodic boundary conditions for all
derivatives of the path up to $h^{[n-1]}$ is understood.  The measure
$\mathcal D_n h(t)$ is defined such that the propagator
$G_{n}(\zd|\zd^0;T)$ of Sec.~\ref{sec:alpha-2n} is a path integral
without extra normalization, and the boundary conditions of the
integral are specified by the arguments of $G_{n}$, i.e.
\begin{eqnarray}
  G_{n}(\zd|\zd^0;T)= \int\hspace{-3pt}\mathcal D_n h(t) \exp\left( - \frac 12
\int_0^T\td t\,  [h^{[n]}(t)]^2 \right)\!,\,
\label{green1}
\end{eqnarray}
with $h^{[n-k]}(0)=z_k^0,\, h^{[n-k]}(T)=z_k$, where $k=1,\dots n$.
This $\mathcal D_n h(t)$ is in fact the measure leading naturally to
the quantum-mechanical-like operator representation of the path
integral
\begin{eqnarray}
  G_{n}(\zd|\zd^0;T)= \left\langle \zd | e^{-\hat H_n^0 T} |
  \zd^0 \right\rangle ,
\end{eqnarray}
where $\hat H_n^0$ is given by Eq.~\eqref{eq:H0}, and $\left|\zd^0
\right\rangle (\left\langle \zd\right|$) are its right (left)
eigenvectors corresponding to the $n$ dimensional positions indicated
therein.  Note that since this Hamiltonian is non-Hermitian for $n\geq
2$, the left and right eigenfunctions are different in general.

A third version of the propagator we shall utilize comes from a
path integral by a measure of one order lower as
\begin{eqnarray}
  G_{n+1}(\zd|\zd^0;T)&=& \int\hspace{-3pt}\mathcal D_n h(t) \exp\left( - \frac
12 \int_0^T\td t\,  [h^{[n]}(t)]^2 \right) \nonumber \\
&& \times \delta\left( z_{n+1}-\int_0^Th(t)\td t -z_{n+1}^0 \right).
\label{eq:Gn+1}\end{eqnarray}
Here the Dirac delta produces the normalized density for the added
area variable $z_n+1$. Note that if we take into account
Eq.~\eqref{green1} then the consistency relation \eqref{eq:GF-int}
immediately follows.

In order to determine the normalization coefficient $\mathcal A$ in
\eqref{eq:mrh-pi}, we express the equal-points propagator complemented
with the area variable set to zero at both ends.  By integrating over
the path except for a single point and using Eqs.~\eqref{eq:mrh-pi}
and \eqref{eq:Gn+1}
\begin{eqnarray}
  P_{\text PBC}(\zd^0)&=&\int_{\text{PBC}:\zds^0} \hspace{-10pt}
  \mathcal D_n h(t) P[h(t)] \nonumber \\
  &=&\mathcal{A} G_{n+1}(\zd^0,0|\zd^0,0;T).
  \label{eq:gf-mrh}
\end{eqnarray}
where the mark PBC:$\zd^0$ refers to the time derivatives at the ends
fixed at $h^{[k]}(0)=h^{[k]}(T)=z_{n-k}^0, \, k=0,\dots,n-1$.  The
$P_{\text PBC}(\zd^0)$ is the joint probability density of
$h^{[k]}(t)=z_{n-k}^0$-s
in a periodic path at any fixed time $t$, so as a byproduct we
obtained that density in terms of the propagator, explicitly given in
Sec.~\ref{sec:alpha-2n}.  That joint probability density is obviously
normalized to unity.  However, we know from Eq.~\eqref{eq:int-per-gf}
that the integral of the r.h.s. is independent of the $z_{n+1}^0$-st
variable, and therefore we have
\begin{eqnarray}
  \mathcal{A} = T^{n+\frac 12} \sqrt{2\pi}.\label{eq:normalizer}
\end {eqnarray}
For $n=1$ the normalizing coefficient derived
in~\cite{MajumdarComtet:2004,MajumdarComtet:2005} is recovered.

The integrated distribution $M(h_\text{m};T)$ of the MRH,
i.e., the probability that the
maximum does not exceed $h_\text{m}$, has been formulated in terms of
a path integral in~\cite{MajumdarComtet:2004,MajumdarComtet:2005}. That
expression is valid for any path density $P[h(t)]$ and reads formally as
\begin{eqnarray}
  M(h_\text{m};T) &=& \int_{-\infty}^{h_\text{m}} \hspace{-5pt} \td
  h\, P(h;T) \nonumber \\ &=&\int_{\text{PBC}}\hspace{-10pt}
  \mathcal D h(t) P[h(t)] \prod_t \theta(h_\text{m}-h(t)).
\end{eqnarray}
Note that here $P(h_{\text m};T)$ is the density of MRH.
Changing the integration variable and then introducing the hard wall
potential $V_0(h)=\infty$ for $h<0$ and $V_0(h)=0$ for $h>0$, one
obtains
\begin{eqnarray}
  M(h_\text{m};T) \hspace{-2pt}&=& \hspace{-5pt}\int_{\text{PBC}}\hspace{-15pt}
\mathcal D h(t)
  P[h_\text{m}-h(t)] \prod_t \theta(h(t))
  \nonumber \\
  \hspace{-2pt}&=& \hspace{-5pt}\int_{\text{PBC}}\hspace{-15pt} \mathcal D h(t)
  P[h_\text{m}-h(t)]  e^{-\int_0^T \hspace{-2pt}\td t\, V_0(h(t))}.
\end{eqnarray}
Using the specific form \eqref{eq:mrh-pi} of the probability
functional we find
\begin{eqnarray}
  M(h_\text{m};T) &=&\mathcal{A} \int_{\text{PBC}}\hspace{-10pt}
  \mathcal D h(t) \, \delta\left(h_\text{m}T-
    \int_0^T\hspace{-5pt}\td t
    \, h(t)\right) \nonumber \\&\times&\exp\left[ -
    \int_0^T\hspace{-5pt} \td t\,\left\lbrace  \frac 12( h^{[n]})^2
      + V_0(h(t))\right\rbrace \right].
\end{eqnarray}
Next, we introduce the scaled Laplace transform of the integrated MRH
distribution
\begin{eqnarray}
  K(u;T) &=& T\int_0^\infty \hspace{-5pt} \td h_\text{m}
  e^{-uh_\text{m}T}
  M(h_\text{m};T) \nonumber \\
  &=& \mathcal{A} \int_{\text{PBC}}\hspace{-10pt} \mathcal D h(t)
  \nonumber \\ &\times&\hspace{-5pt} \exp\left[ - \int_0^T\hspace{-5pt}\td
    t\,\left\lbrace\frac 12( h^{[n]})^2 + uV(h(t))\right\rbrace \right],
\label{eq:K}\end{eqnarray}
where the potential $V(h)=\infty$ for $h<0$ and $V(h)=h$ for $h>0$.

In order to find $K(u;T)$, we write down the evolution equation for
the PDF of the position and its derivatives $P_n(\zd;t)$ corresponding
to the above path probability
\begin{eqnarray}
  \partial_t P_n  &=& - \hat H_n(u) P_n \label{eq:FP-n},\\
  \hat H_n(u) &=& \hat H_n^0 + uV(z_n),\label{eq:H}
\end{eqnarray}
where $\hat H_n^0$ was given in \eqref{eq:H0} and the variables $z_k$
are defined by \eqref{eq:z-x}.  Thus the Laplace transform can be
written in short as
\begin{eqnarray}
  K(u;T) = \mathcal{A}\, \text{Tr} \exp\left( -\hat H_n(u)T\right).
\end{eqnarray}

It is straightforward to show that the eigenvalues $E_{n,\omega}(u)$
of $\hat H_n(u)$, where $\omega$ summarizes all discrete indices, obey
a simple scaling in $u$.  For that purpose, let us consider the
eigenvalue problem for $h=z_n>0$
\begin{eqnarray}
\hat H_n(u)\psi =\hspace{-3pt}
\left[-\frac 12 \partial_1^2 +\sum_{k=1}^{n-1} z_k
  \partial_{k+1} + u z_n\right] \psi = E_n(u) \psi
\end{eqnarray}
and apply a scale transformation by substituting
\begin{eqnarray}
u^{\beta_k}z_k \to z_k,\ \ \ u^\delta E_n\to E_n.
\end{eqnarray}
We recover an equation free of $u$, if all powers multiplying various
terms are the same, that is
\begin{eqnarray}
-2\beta_1=\beta_1-\beta_2=\dots =\beta_{n-1}-\beta_n=\beta_n+1=\delta,
\end{eqnarray}
so $\beta_k=(n-k+1)\delta-1$ and also $-2\beta_1=\delta$.  Hence
$\delta = \frac{2}{2n+1}$, so the eigenvalues scale like
$E_{n,\omega}(u) = \epsilon_{n,\omega}u^\frac{2}{2n+1}$, where
$\epsilon_{n,\omega}$ is the spectrum of $\hat H_n=\hat H_n(1)$.

It thus follows that, using \eqref{eq:normalizer}, we get the scaling
relation for the Laplace transform of the integrated distribution
\begin{eqnarray}
  K(u;T) &=& T^{n+\frac 12} {\cal K}(uT^{n+\frac 12}) \\
  {\cal K}(s)&=&\sqrt{2\pi} \, \text{Tr} \exp\left( -\hat H_n
    s^\frac{2}{2n+1} \right)\nonumber \\
  &=&\sqrt{2\pi}  \,\sum_\omega \exp\left( -\epsilon_{n,\omega}
  s^\frac{2}{2n+1}\right).
\end{eqnarray}
Hence, using \eqref{eq:K}, we obtain for the PDF of the MRH
$P(h_\text{m};T)$ and its moment generating function $G(u;T)$ in
scaling forms
\begin{eqnarray}
  P(h_\text{m};T) &=& \partial_{h_\text{m}} M(h_\text{m};T) =
  T^{\frac 12 -n} P(h_\text{m}T^{\frac 12 -n}), \label{scaledMRH}\\
  G(v;T) &=& \int_0^\infty\hspace{-5pt} \td h_\text{m}\,
  P(h_\text{m};T) e^{-vh_\text{m}} = G(vT^{n-\frac 12 }),\phantom{xx}
\end{eqnarray}
where
\begin{eqnarray}
  G(s)&=& \int_0^\infty\td z\, P(z) e^{-sz}\nonumber \\
  &=&s\,{\cal K}(s) \nonumber \\
  &=& \sqrt{2\pi} \,s \,\text{Tr} \exp\left( -\hat H_n
    s^\frac{2}{2n+1}\right)
  \nonumber \\ &=& \sqrt{2\pi} \,s \,\sum_\omega \exp\left( -\epsilon_{n,\omega}
    s^\frac{2}{2n+1}\right). \label{eq:gen-trace}
\end{eqnarray}
Note that the same symbols $P,G$ are used for single- and
double-argument functions, but that should not cause confusion.
Remarkably, the trace formula is exactly the same as in the case
of the simple random walk $n=1$, with the Hamiltonian $\hat H_1$
replaced by $\hat H_n$. Note that, in the special case of $n=1$,
the scaling function of the
MRH distribution \eqref{Majumdar-full} is ultimately recovered from
the above trace formula \cite{MajumdarComtet:2005}.

The scaled moment generating function Eq.~\eqref{eq:gen-trace}
together with the preceding scaling formulas are our main result
here. In the next two sections, we shall exploit the above results to
draw conclusions about the scale and the small argument asymptote of
the MRH distribution function.

\section{Strong-correlation regime ($1<\alpha <\infty$)}
\label{alpha.gt.1}

\begin{figure}[htb]
  \includegraphics{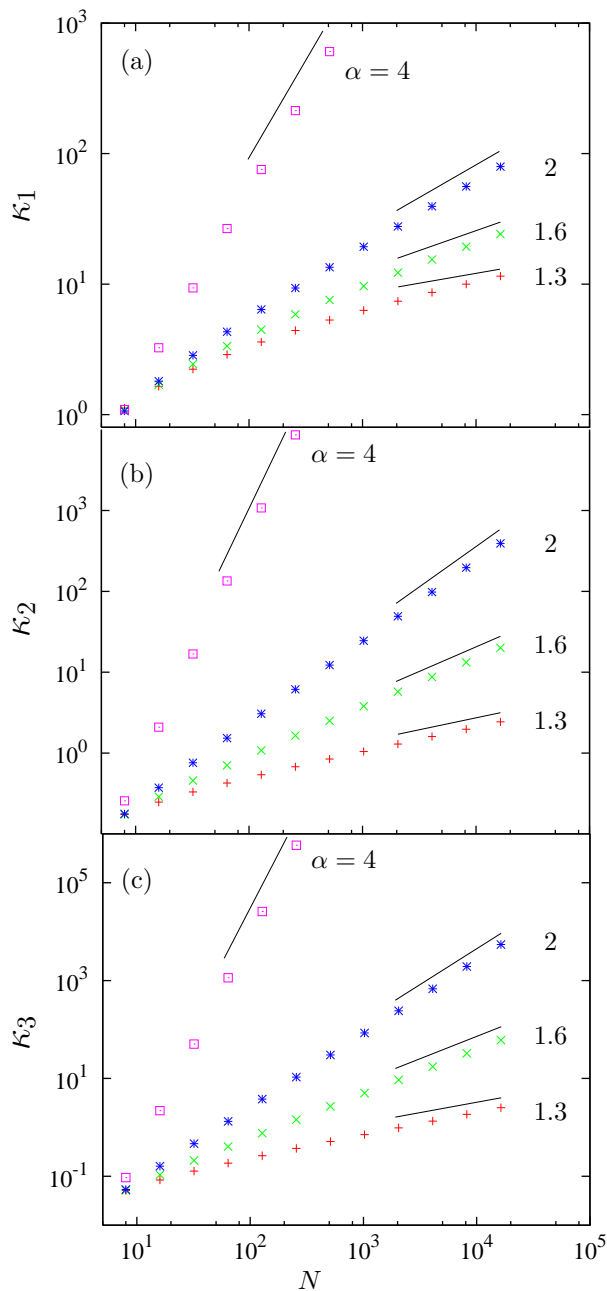}
\caption{Cumulants, $\kappa_k$, of the MRH distribution for various
$\alpha>1$ showing that $\kappa_k(\alpha)$ scales with system size,
$N=T/\tau$, as $\kappa_k(\alpha)\sim N ^{k(\alpha -1)/2}$. The
straight lines have the appropriate asymptotic slopes $k(\alpha
-1)/2$.
\label{F:cumulants}}
\end{figure}

In order to evaluate the trace formula one would need the energy
eigenvalues of $\hat H_n(1)$. Although they are
known~\cite{MajumdarComtet:2004} only for $n=1$, assuming that these
eigenvalues exist, the scale of the MRH in $T$ can be derived since
Eq.~\eqref{scaledMRH} yields
\begin{equation}
\langle h_m\rangle \sim T^{\, n-\frac{1}{2}} \quad .
\label{h_maverage}
\end{equation}
As one can see, the scale of $\langle h_m\rangle$ is the same as that
of the square root of the roughness~\cite{AntalETAL:2002}, i.e.  we
have $\langle h_m\rangle\sim \sqrt{\langle w_2\rangle }$ just as in
the case of random walks~\cite{MajumdarComtet:2004}.  It should be
emphasized that while the above reasoning holds strictly for
$\alpha=2n$, the exponent can, in fact, be continued naturally to real
values.  Thus it is plausible to surmise that $T^{\frac{\alpha-1}{2}}$
is the scale of the MRH for any $\alpha>1$.  This power emerges quite
sharply for $\alpha=1.3, 1.6, 2,4$ in numerical simulations as shown
on the first panel of Fig.~\ref{F:cumulants}.

Fig.\ \ref{F:cumulants} also displays the 2nd and 3rd cumulants of
$P(h_m,T)$.  We can observe the emergence of well defined scaling with
$T$
\begin{equation}
\kappa_k(\alpha)\sim T ^{k(\alpha -1)/2} \quad .
\label{cumulants-sim}
\end{equation}
The scaling exponents are again equal to those of the cumulants of the
width-distribution provided the $h_m\sim \sqrt{w_2}$ correspondence is
used. This suggest that there is an intimate connection between the
fluctuations of MRH and those of the signal width.

\begin{figure}[htb]
{\includegraphics{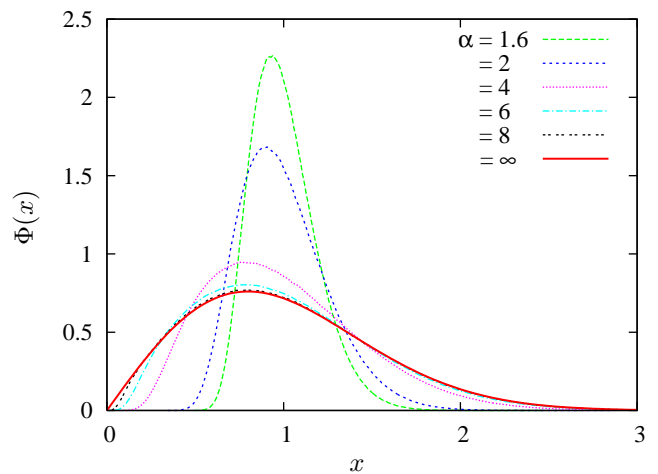}}
\caption{Numerically constructed MRH distributions for various
 $1<\alpha<\infty $.  The $\alpha =\infty$ curve is the analytic
 result given in Eq.~\eqref{a-inf-pdf}. System sizes $N\le 16384$ were
 sufficiently large to observe the convergence of the PDF-s within the
 width of the lines drawn. Note that the $\alpha =8$ results are
 almost indistinguishable form the $\alpha\to\infty$ limit.}
\label{F:picture_gallery}
\end{figure}

In order to see how the general shape changes as $\alpha$ is
increased, we have performed simulations as described in
Sec.~\ref{technicalities}.  The results are shown in
Fig.\ \ref{F:picture_gallery} where we used {\em scaling by the average}
to present the scaling functions $\Phi_\alpha (x)$.

The main features can be readily seen.  The scaling function is a
unimodal (single peaked) function which spreads out as $\alpha$ increases and
approaches its $\alpha\to\infty$ limit (see Eq.~\eqref{a-inf-pdf})
rather fast. This is not entirely surprising since a glance at
Fig.\ \ref{F:alpha_scan} convinces one that the $\alpha =10$ signal
already consists of a single mode for all practical purposes, and thus
the MRH distribution will be very well approximated
by the $\Phi_\infty (x)$ function.

The function decays to zero extremly fast in the $x\to 0$ limit.  The
nonanalytical behavior and the actual functional form at small $x$
will be the subject of the next Section. Here, we call the reader's
attention to the fact that the region where the nonanalytic asymptotic
behavior dominates is shrinking as $\alpha$ increases and, according to
Eq.~\eqref{a-inf-pdf}, entirely disappears in the $\alpha\to \infty$
limit.

The large $x$ limit is harder to treat analytically and we have only
numerical evidence (Fig.\ \ref{F:smallx}) that the asymptotic behavior
for large $x$ is given by
\begin{equation}
\Phi (x)\sim Cx^\gamma e^{-Bx^2}
\label{large-x-asymptote}
\end{equation}
where the parameters $B$, $C$, and $\gamma$ depend on $\alpha$.  The
above functional form is consistent with the exact result at $\alpha
=\infty$ (see Eq. \eqref{a-inf-pdf}). At $\alpha =2$, the ansatz of a
Gaussian decay was shown to be in
agreement~\cite{MajumdarComtet:2004} with the large-order moments of
the distribution function. However, the possibility of a prefactor
$x^\gamma$ was not excluded by the analysis.  We found that the
generalized asymptote \eqref{large-x-asymptote} with $\gamma\approx 2$
gives a superior fit to the large-$x$ ($x>1.5$) behavior of the
exactly known PDF.

We have also fitted our numerical data in the region $x>1.5$ for
larger $\alpha$-s, resulting in $\gamma\approx 1.4$ and $\gamma\approx
1.1$ for $\alpha =3$ and $4$, respectively.  The general trend of the
exponent $\gamma$ with increasing $\alpha$ is consistent with the
$\alpha\to \infty$ limit of $\gamma_\infty=1$.

\section{Initial asymptote}
\label{sec:small-x}

The trace formula \eqref{eq:gen-trace} allows us to perform an
asymptotic analysis of the MRH distribution for small arguments.  The
calculation is based on the large $s$ behavior of the moment
generating function \eqref{eq:gen-trace}, wherein we assume that there
is a positive, $\alpha=2n$-dependent, nondegenerate ground state
energy $\epsilon_0(\alpha)$, which gives the leading term of the sum
(while $\alpha$ is strictly even, several results will lend themselves
to continuation). Under this assumption, the PDF in the scaled
variable $z=h_\text{m}T^{\frac{1-\alpha}{2}}$ is asymptotically given
by
\begin{eqnarray}
  P(z)&=& \int  \frac{\td s}{2\pi i} G(s) e^{sz} \nonumber \\
  &\approx& \int  \frac{\td s}{2\pi i} s \sqrt{2\pi}\, \exp\left(
    sz-\epsilon_0 s^{\frac{2}{\alpha+1}} \right).
  \label{appeq:pdf-mrh}\end{eqnarray}
The above integral can be calculated using the saddle point method.
For small $z$, the saddle point of the exponent is located
on the real axis at
\begin{eqnarray}
  s^\ast\approx \left(\frac{2\epsilon_0}{(\alpha+1)z}
  \right)^{\frac{ \alpha+1}{\alpha-1}}
\end{eqnarray}
and the integral in the neighborhood of the saddle point reduces to
evaluating a Gaussian integral which yields the following asymptote
\begin{eqnarray}
  P(z) &\approx& C z^{-\gamma} \exp \left( - B/z^\beta\right),
  \label{eq:asymp-pdf-mrh}
\end{eqnarray}
where the parameters are given by
\begin{subequations}
\begin{eqnarray}
  \beta &=& \frac{2}{\alpha-1},\\
  \gamma &= &\frac{2\alpha+1}{\alpha-1},\\
  B &= & \frac{\alpha-1}{2} \left(\frac{2\epsilon_0}{\alpha+1}
  \right)^{\frac{\alpha+1}{\alpha-1}}\\
  C &= &  \sqrt{\frac{\alpha+1}{\alpha-1}}
  \left(\frac{2\epsilon_0}{\alpha+1} \right)^{\frac 32
    \frac{\alpha+1}{\alpha-1}}.
\end{eqnarray}
  \label{eq:asymp-parms-mrh}
\end{subequations}
One should note that the exponents $\beta$ and $\gamma$ depend only on
$\alpha$ while the amplitudes also depend on the ground state energy,
$\epsilon_0(\alpha)$. The value of $\epsilon_0(\alpha)$ is known only
for $\alpha=2$ where $\epsilon_0(2)=\alpha_1/\sqrt[3]{2}$, with
$\alpha_1=2.3381$ being the absolute value of the first zero of the
Airy function~\cite{MajumdarComtet:2004,MajumdarComtet:2005}.
\begin{figure}[htb!]
\includegraphics{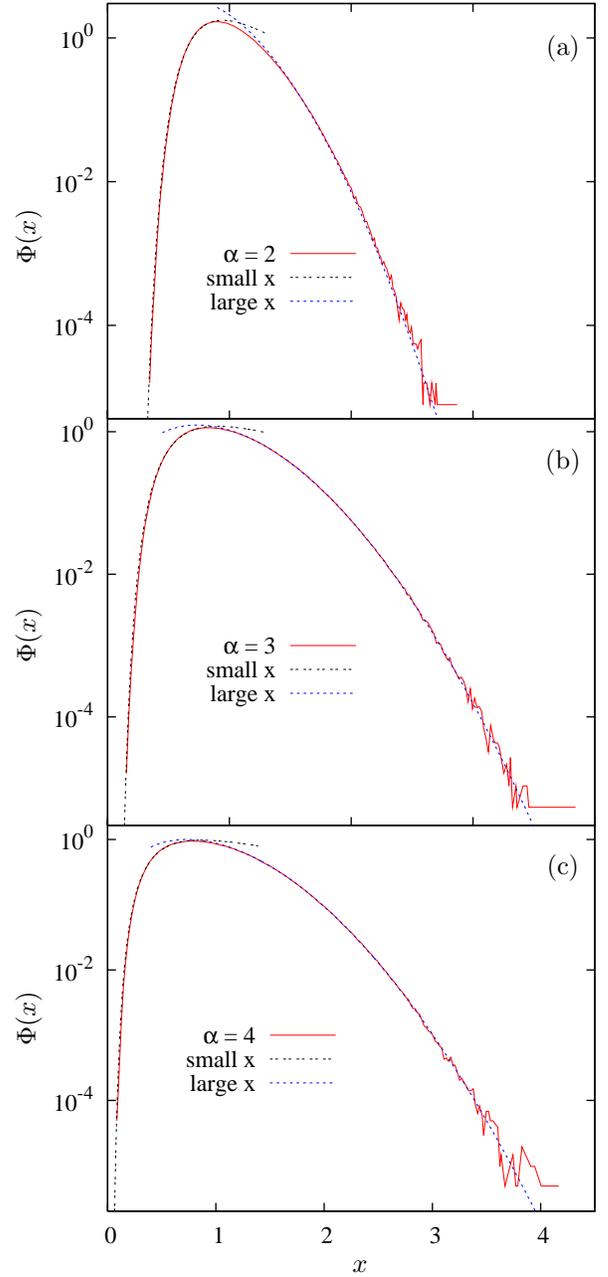}
\caption{MRH distributions for (a): $\alpha=2$, (b): $\alpha=3$, and
  (c): $\alpha=4$, calculated for system size $N=16384$ (solid lines).
  The small and large $x$ asymptotes (dashed lines) are also shown.
  The small $x$ behavior in the range $0<x<1$ is fitted to the
  functional form \eqref{eq:asymp-pdf-mrh} with the exponents
  $\beta,\gamma$ taken from \eqref{eq:asymp-parms-mrh}. The prefactors
  $B,C$ are fitting parameters (note that the formulas in
  \eqref{eq:asymp-parms-mrh} contain an unkown parameter
  $\epsilon_0$). Large $x$ data in the range $x>1.5$ are fitted to the
  form \eqref{large-x-asymptote}, where $B$, $C$ and $\gamma$ are
  fitting parameters.
  \label{F:smallx}}
\end{figure}

It should be emphasized that we did not scale the mean to $1$, being
ignorant about the full PDF as well as its mean for general $\alpha$.
So if comparing the above formula to the numerically scaled PDF as
function of $x=z/\left\langle z\right\rangle$ then the factors $B,C$
will change and become fitting parameters.  Figure \ref{F:smallx}
demonstrates the fit of \eqref{eq:asymp-pdf-mrh} to simulation results
for several $\alpha$-s, and we find that the fits are excellent in a
surprisingly large interval.  It should be noted that in the large
$\alpha$ limit the initial slope is positive, so one expects a
decreasing range of validity of the asymptote for increasing $\alpha$,
nevertheless, the fit on Fig.~\ref{F:smallx} is quite good even for
the largest $\alpha$.  The case $\alpha=3$ demonstrates the
continuation of the $\alpha=2n$ based formula, and suggests that naive
continuation of at least the exponents $\beta,\gamma$ in
\eqref{eq:asymp-parms-mrh} is justified.

Returning to the problem of scale-dependence of the amplitudes $B$ and
$C$, we note that even if the full PDF is unknown, one can construct a
parameter from the small-$x$ asymptote which does not depend on the
scale.  In order to see this, let us consider scaling by the
average. With the rescaled variable $x$, one has the PDF as
$\left\langle z\right\rangle P(x\left\langle z\right\rangle)$ and
writing it again in the form \eqref{eq:asymp-pdf-mrh} yields the
following change of the amplitudes
\begin{eqnarray} B' = \frac{B}{\left\langle z\right\rangle^\beta}, \ \
  \ C' = \frac{C}{\left\langle z\right\rangle^{\gamma-1}} \, .
\end{eqnarray}
It follows from the above expressions that the following combination
\begin{eqnarray}
  D&=&\frac{B^{\frac{\alpha+2}{2}}}{C}=
  \frac{B'^{\frac{\alpha+2}{2}}}{C'}
  \nonumber \\
  &=& \frac{(\alpha-1)^{\frac{\alpha+3}{2}}}{2^{\frac{\alpha+2}{2}}
    \sqrt{\alpha+1}} \left(\frac{2\epsilon_0}{\alpha+1}
  \right)^{\frac{\alpha+1}{2}}
\end{eqnarray}
remains independent of any scale change.

We should reiterate that the energy parameter $\epsilon_0(\alpha)$ is
not known generally, but it is plausible to assume that it is a well
defined number.  It may be determined numerically for $\alpha=2n$ by a
direct study of the corresponding local Hamiltonian.  Remarkably,
however, the above asymptotic formula allows for the computation of
$\epsilon_0(\alpha)$ for any $\alpha>1$ from a numerical fit of the
simulation result. Thus, precise MRH statistics effectively extract
the ground state energy level of the Hamiltonian without solving the
corresponding differential equation.  Continuation of
\eqref{eq:asymp-pdf-mrh} for $\alpha\neq 2n$ is also natural here, but
in this case we have a non-local Hamiltonian, whose spectral problem
would be an even more challenging task to solve. Unfortunately, very
high precision simulations are required to determine the ground state
energy from the small-$x$ asymptote. In particular, our simulated data
did not even allow the computation of the ground state energy to
within a factor of 2 for the case of $\alpha =2$ where the lowest
eigenvalue is known.

\section{MRH distribution for large $\alpha$}
\label{sec:large-alpha}

We have calculated the MRH distribution for the $\alpha\to\infty$
limit in Sec.~\ref{S:exact-results}. There we found that only the $n=1$
mode survives and, as a result, the PDF
\eqref{a-inf-pdf} emerges.  Here we discuss a procedure for
perturbatively computing the leading corrections to \eqref{a-inf-pdf}
by taking into account the modes $n=2,3,\dots$.

First we reiterate that the amplitude of modes $c_n$ obey the
distribution with action \eqref{E:fourier_action} and measure
proportional to $\prod_n \theta(c_n) c_n\mathrm{d}c_n$.  Thus,
separating the $n=1$ mode, the path in Fourier representation is
written as
\begin{equation}
  h(t)= a_1 \sin(t) + \sum_{n=2}^\infty \varepsilon_n a_n \sin(nt+\varphi_n),
\end{equation}
where the $\varepsilon_n=1/n^{\alpha/2}$ is the mean square root
deviation of the amplitude of the $n$-th mode, and the
$a_n=c_n/\varepsilon_n$-s are i.i.d.\ variables distributed according
to
\begin{equation}
  P_0(z)= 2z\theta(z)\exp(-z^2) \, .
\label{WD-pdf}
\end{equation}
Finally the phases $\varphi_n$ are independent and uniformly
distributed in $[0,2\pi]$.  The $n=0$ phase is omitted, because the
choice of the origin is arbitrary.  Obviously, $x$ measures the height
from the time average of the path, which is here set to zero. Note
that now time $t$ is in units of $2\pi/T$.

The leading correction from higher frequency modes can be calculated
independently for each mode, thus here we only consider the $n$-th
mode.  Then the path is
\begin{equation}
  h(t)= a_1 \sin(t) + \varepsilon_n a_n \sin(nt+\varphi_n)
\label{eq:path-two-modes}
\end{equation}
and the calculation to leading order is straightforward. We compute
the maximum of the path and then, knowing the distribution of all
parameters therein, we can determine the PDF of the maximum.  The
details are presented in Appendix \ref{app:large-alpha}, where we
obtain the perturbed PDF for $h_mT^\frac{1-\alpha}{2}=z$ as
\begin{equation}
  P(z) = P_0(z)+ \varepsilon_n^2 P_{2,n}(z)
\label{eq:pert-pdf-unsc-1}
\end{equation}
with
\begin{eqnarray}
  P_{2,n}(z) &=&  (1-n^2/2)\delta(z) +z \delta'(z)/2 \nonumber \\
   &&+e^{-z^2} \theta(z) \left( 2z^3 + (n^2-3) z\right).
  \label{eq:pert-pdf-unsc-2}
\end{eqnarray}
The singular part needs some explanation here. As has been discussed
in Sec.~\ref{sec:small-x}, for finite $\alpha$-s the PDF starts
nonanalytically with zero initial slope for finite $\alpha$-s, in
contrast to the $\alpha=\infty$ case, where the PDF has a finite
slope.  The nonanalyticity is not expected to be recovered by any
expansion.  Nonetheless, the formal expansion gives an explicit
correction function $P_{2,n}$, with delta-singularity at the origin.
It is plausible to conclude that while the expansion cannot be
correct overall, the singularity ``tries" to take care of the nonanalytic
difference in the small-$z$ behavior, while the nonsingular part is
expected to be a faithful correction for $z>0$.  This leaves open the
possibility that the large $\alpha$ expansion is not convergent,
rather it is asymptotic.

\begin{figure}
\includegraphics{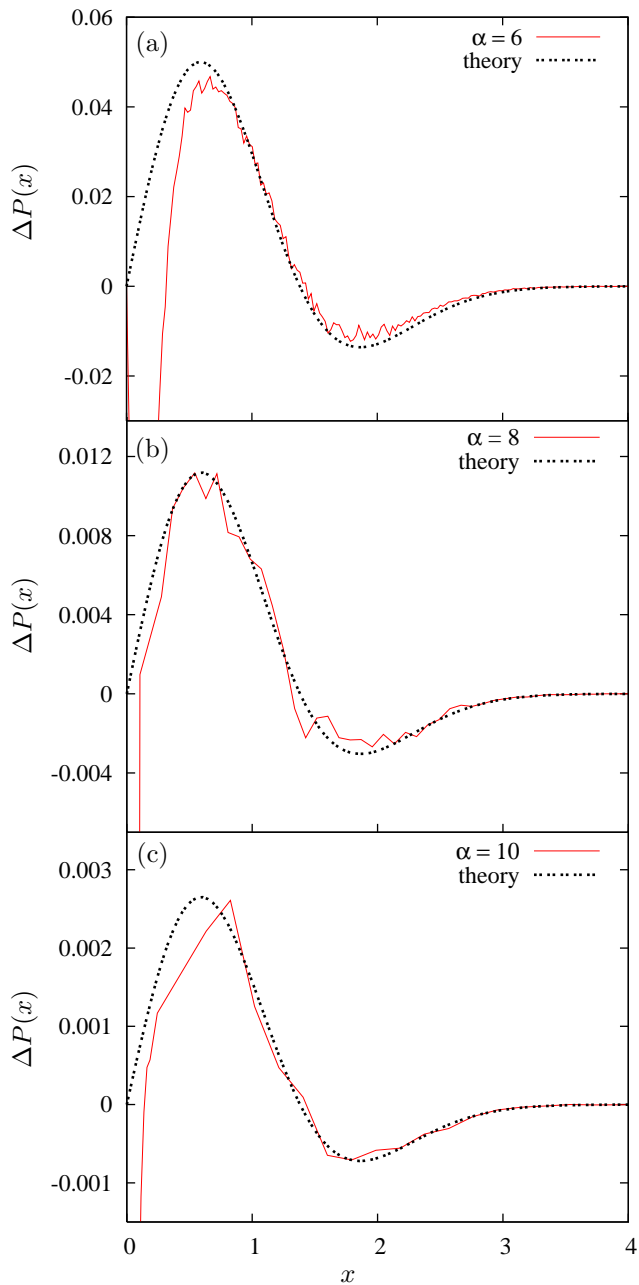}
\caption{Difference $\Delta P(x)$ between the MRH distribution for
  large $\alpha$ and $\alpha=\infty$, as defined in
  \eqref{eq:delta-P}.  We also display the smooth part of the leading
  correction \eqref{eq:delta-P23} from perturbation theory added up
  from the modes $n=2,3$.
  \label{F:alpha_infinity}}
\end{figure}

Next we scale the PDF to unit average.  Using the result
(\ref{appeq:pert-pdf-mean})
from Appendix \ref{app:large-alpha} one finds
\begin{equation}
  \Phi (x) = \Phi_{_\infty}(x)+ \varepsilon_n^2 \Phi_{2,n}(x)
  \label{eq:pert-pdf-sc-1}
\end{equation}
where $\Phi_{_\infty}(x)$ is given in \eqref{a-inf-pdf} and
\begin{eqnarray}
  \Phi_{2,n}(x) &=&  (1-n^2/2)\delta(x) +x \delta'(x)/2\nonumber \\
  && +e^{-\pi x^2/4} \theta(x) \frac{\pi}{8}(n^2-1)\left( 6 x -\pi x^3\right).
  \label{eq:pert-pdf-sc-2}
\end{eqnarray}
Formally, we can sum up the leading corrections for all $n$-s.
However, this is not a consistent approximation, because, for
instance, $\varepsilon_2^2=\varepsilon_4$, so the leading correction
from $n=4$ is of the same order as the quadratic one from $n=2$.
Therefore we use the sum of the corrections for only $n=2,3$ to test the
prediction.  On Fig.\ \ref{F:alpha_infinity} we display the correction
\begin{eqnarray}
\Delta \Phi(x)=\Phi (x)-\Phi_{_\infty}(x)
  \label{eq:delta-P}
\end{eqnarray}
of the PDF from simulation for a series of $\alpha$-s, together with
the theoretical prediction for corrections added up from the modes
$n=2,3$,
\begin{eqnarray}
\Delta \Phi (x)\approx
\varepsilon_2^2 \Phi_{2,2}(x)+ \varepsilon_3^2 \Phi_{2,3}(x).
  \label{eq:delta-P23}
\end{eqnarray}

It should be mentioned that the sum of leading corrections for all
modes has a prefactor
$\sum_{n=2}^\infty(n^2-1)\varepsilon_n^2=\zeta(\alpha-2)-\zeta(\alpha)$.
This diverges for $\alpha\to 3^+$, so the series does not converge for
$\alpha \le \alpha_c=3$, which is the borderline for the
differentiability of the path.  We cannot exclude that higher order
corrections, involving higher order differentiation of the path,
further tighten the range of convergence.

\section{Comparison with the roughness and the maximal intensity}
\label{sec:comp-rough}

Here we compare the statistical properties of $h_m$ to those of the
roughness $w_2$, i.e.  of the mean square deviation, or width of the
trajectory \eqref{w2-eq}.  The latter was one of the first global quantities of
stochastic signals whose scaling properties and statistics were
extensively studied for $1/f^\alpha$ processes~\cite{AntalETAL:2002}.

One of the reasons for comparison is that both $\langle h_m \rangle$
and $\sqrt{\langle w_2 \rangle}$ scale similarly with $T$ for
$\alpha \ge 1$ and,
furthermore, there are many common features at the level of their
PDF-s. Namely, for algebraically diverging correlations, $\alpha> 1$,
after scaling by the mean, the PDF-s are nondegenerate (each cumulant
is finite), that is, scaling by the mean is a natural representation
of both PDF-s.  As $\alpha\to 1$, the PDF-s scaled by the mean
approach a Dirac-delta and, in the range $\alpha\leq 1$, they both
lend themselves to scaling by the standard deviation.  Here an
important difference emerges. In the range $0.5< \alpha\le 1$ the
roughness has a nontrivial PDF while below the critical $\alpha=0.5$
it becomes trivial, i.e. the roughness becomes Gaussian
distributed. On the other hand, the MRH has the trivial FTG limiting
distribution in the entire $0<\alpha<1$ region. We can only speculate
that the threshold near $\alpha=0.5$ manifests itself in the MRH
distribution in its approach to the FTG limit, as suggested by the
finite-size dependence of the simulation results shown in
Fig.~\ref{F:skewness}. This, however, is just a numerical observation
without theoretical foundations as yet.

Further motivation for a closer comparison comes from the similarities
in the shape of the two families of PDF-s in the $\alpha>1$ region.
First, in the $\alpha\to\infty$ limit, the PDF-s are the same if the
$h_m\sim \sqrt{w_2}$ correspondence is made.  Second, for finite
$\alpha$-s, the unimodal PDF-s have asymptotes which are similar for
both small and large arguments.  Specifically, there is a Gaussian
decay at large $x$, while the small $x$ behavior is dominated by an
exponential nonanalytic term with a power prefactor. Here, the
comparisons can be made quantitatively for small $x$, since analytic
results are available for general $\alpha$.

Last but not least, a reason for a closer comparison comes from the
fact that the roughness can also be conceived as obeying an
EVS. Bertin and Clusel~\cite{Bertin:2005,BertinClusel:2006} made the
remarkable observation that since the roughness is essentially the
integrated power spectrum, i.e., the sum of nonnegative Fourier
intensities, it is in effect the maximum of positive partial sums. In
general, the partial sums are correlated but, for the special case of
$\alpha=1$, they correspond to the ordered sequence of
i.i.d.\ variables. As a consequence,
FTG distribution emerges for $w_2$ at $\alpha=1$, thus providing
insight to an
earlier rather puzzling result~\cite{AntalETAL:2001} in
connection with $1/f$ noise. It then becomes a rather interesting open
question how the MRH distribution differs from the roughness
distribution for $\alpha >1$ where the latter also describes the EVS
of correlated variables.

The initial asymptote of the MRH distribution
\eqref{eq:asymp-pdf-mrh},\eqref{eq:asymp-parms-mrh} should be compared
with that for the roughness distribution obtained in appendix E
of~\cite{GyorgyiETAL:2003}
\begin{eqnarray}
  \Phi_w(x) &\approx& C_w x^{-\gamma_w} \exp \left( - B_w/x^{\beta_w}\right),
  \label{w2-small-x-general}
\end{eqnarray}
where the parameters are given by
\begin{subequations}
\begin{eqnarray}
  \beta_w &=& \frac{1}{\alpha-1},\\
  \gamma_w &= &\frac{3\alpha-1}{2(\alpha-1)},\\
  B_w &= & (\alpha-1) \left( \frac{\pi}{\alpha \sin(\pi/\alpha)}
  \right)^{\frac{\alpha}{\alpha-1}} \zeta(\alpha)^{-\frac{1}{\alpha-1}} \\
  C_w &= & \frac{(2\pi)^{\frac{\alpha-1}{2}}}{\sqrt{\alpha-1}}
  \left(\frac{\pi}{\sin(\pi/\alpha)}\right)^{\frac{\alpha}{\alpha-1}}
  \left(\zeta(\alpha)\alpha \right)^{-\frac{\alpha+1}{2(\alpha-1)}}
  \label{w2-small-x-details}
\end{eqnarray}
\end{subequations}
with $\zeta(\alpha)$ denoting the Riemann's zeta function.  Note that
this asymptote does not contain unknown parameters such as
$\epsilon_0$ in the MRH distribution.

Interestingly, comparison with the exponents in the asymptote of the
MRH, \eqref{eq:asymp-pdf-mrh}, shows that the respective $\gamma$-s
are the same, if $\sqrt{w_2}$ is considered, i.e., $2\gamma_w=\gamma$.
Nevertheless, the respective exponents in the prefactor,
$2\beta_w+1$ and $\beta$, agree only at an accidental $\alpha$ and are
otherwise different.

The present results on the small-$x$ asymptote may also be compared to
the asymptote of the distribution of the maximal Fourier intensity.
It is defined as the maximal of the $|c_n|^2$ intensity components for
a given realization of the path, which obeys some PDF if the ensemble
of $1/f^\alpha$ paths is considered.  This was to our knowledge the
first quantity whose EVS was studied in the context of $1/f^\alpha$
signals~\cite{GyorgyiETAL:2003}.  Again, the overall shape of the PDF
of the extremal intensity is similar to those of the MRH and the
roughness: its initial part is suppressed nonanalytically and has a
single maximum, before smoothly decaying for large arguments.  There
the critical $\alpha_c$ where the FTG limit distribution emerges is
$\alpha_c=0$, in contrast to the MRH and the roughness, where this
critical values are $\alpha_c=1$ and $\alpha_c=0.5$, respectively.
As we have shown
in~\cite{GyorgyiETAL:2003}, written with $\alpha_c$, the powers in the
asymptotic formula for the maximal intensity and for the roughness are
the same
\begin{eqnarray}
  \beta_w=\beta_\text{I}=    \frac{1}{\alpha-\alpha_c},\ \ \
  \gamma_w=\gamma_\text{I}=\frac{3(\alpha-\alpha_c) +2}{2(\alpha-\alpha_c)},
\end{eqnarray}
where the exponents $\beta_\text{I},\gamma_\text{I}$ are defined in
the same way for the initial asymptote of the PDF of the maximal
intensity as $\beta_w,\gamma_w$ were for the PDF of the roughness.

In conclusion, the respective PDF-s of the MRH, the maximal intensity,
and the roughness are similarly looking, unimodal functions, with
nonanalytically slow initial behavior.  Despite the qualitative
similarities, however, it is clear that the three PDF-s are
quantitatively different.  This is natural since they describe
different physical quantities. One may, however, speculate that the
similar features have their roots in the divergent correlations
present in the $\alpha \ge 1$ region.

\section{Final remarks}
\label{summary}

It should be emphasized that we are only at the first stages of
understanding the effects of correlations on EVS. One of the important
tasks for future studies should be the understanding of the
convergence properties in the $0<\alpha <1$ range. Although the limit
distribution is known here, the convergence is extremely slow.  Since
most of the environmental time series of general interest (data on
temperature, precipitation, etc.) correspond to this range, as they
exhibit generically correlations with power-type decay, and the length
of the series is naturally restricted, the development of a theory of
finite-size corrections is important.  The much discussed $\alpha \to
1$ case is even more challenging since it appears to be outside the
reach of present computing abilities.  Thus new analytical approaches
and ideas for numerical recipes are called for.

Another relevant problem is the question of boundary conditions.  It
is known from the $\alpha =2$ case, where both periodic and free
boundary conditions were investigated~\cite{MajumdarComtet:2004}, that
the MRH distribution depends on boundary conditions.  Since the
analysis of a real time series usually means cutting it up into smaller
pieces and making statistics out of the properties of these
subsequences, the appropriate boundary conditions in this case are the
so-called window boundary conditions, when the window under consideration
is embedded in a longer signal.  These boundary conditions have
been discussed in connection with the roughness distribution of
$1/f^\alpha$ signals~\cite{AntalETAL:2002}. It has been found that the
limit distributions depends on the window size (even in the limit of
large external system) and furthermore, the effects become stronger as
$\alpha$ increases. Clearly, similar studies should be carried out for
the EVS problem.

Finally, it remains to be seen if the investigations of the effects of
correlations, in particular the effects of strong correlations, will
allow us a universal classification of EVS similar to that existing
for thermodynamic critical points.

\begin{acknowledgments}
  This research has been partly supported by the
Hungarian Academy of Sciences
(Grants No.\ OTKA T043734 and TS 044839).
NRM gratefully acknowledges support from the
EU under a Marie Curie Intra European Fellowship.
\end{acknowledgments}

\appendix

\section{Derivation of the propagator for $\alpha=2n$}
\label{app:gf-background}

Here we show that the ansatz \eqref{eq:gf-rec} indeed satisfies the
Fokker-Planck equation \eqref{eq:H0} with the coefficients
\eqref{eq:gf-parms}.  Let us start out from \eqref{eq:H0}
\begin{eqnarray}
  \partial_t G_n  = \frac 12 \partial_1^2  G_n -\sum_{k=1}^{n-1} z_k
  \partial_{k+1} G_n,
\label{appeq:gf-fp}
\end{eqnarray}
and substitute
\begin{eqnarray}
G_n=G_{n-1} \mathcal{G},
\end{eqnarray}
where the arguments are understood as in \eqref{eq:gf-rec}. Using the
fact that $G_{n-1}$ also satisfies \eqref{appeq:gf-fp}, we arrive at
\begin{eqnarray}
  \partial_t \mathcal{G} = \frac 12 \partial_1^2 \mathcal{G} +
  \partial_1 \mathcal{G}\,  \partial_1  \ln
  G_{n-1} - \sum_{k=1}^{n-1} z_k
  \partial_{k+1}\mathcal{G}.
\label{appeq:gf-fp-2}
\end{eqnarray}
From \eqref{eq:gf-rec} we have
\begin{subequations}
\begin{eqnarray}
\partial_{k}\mathcal{G} &=& a_k^n \mathcal{G}',\\
\partial_{k}^2\mathcal{G} &=& (a_k^n)^2 \mathcal{G}'',
\end{eqnarray}
\label{appeq:partial-G}
\end{subequations}
where, denoting $\ad^n \!\!\cdot\! \zd - \ad^{0,n} \!\!\cdot\!
\zd^{0}$ by $x$,
\begin{subequations}
\begin{eqnarray}
\mathcal{G}'(x;\sigma) &=& -\frac{x}{\sigma^2} \mathcal{G}(x;\sigma), \\
\mathcal{G}''(x;\sigma) &=& - \frac {\mathcal{G}(x;\sigma)}{\sigma^2} -
\frac {x}{\sigma^2} \mathcal{G}'(x;\sigma),
\end{eqnarray}
\label{appeq:diff-G}
\end{subequations}
and, furthermore, using the full exponent of the ansatz
\eqref{eq:gf-ansatz} we have
\begin{eqnarray}
  \partial_1  \ln G_{n-1} = -  \sum_{k=1}^{n-1}
  \frac{a_1^k}{\sigma^2_k} (\ad^k \!\!\cdot\! \zd - \ad^{0,k} \!\!\cdot\!
\zd^{0}).
\end{eqnarray}
Differentiation of the Gaussian $\mathcal{G}$ by time gives
\begin{eqnarray}
\partial_t \mathcal{G} &=& - \frac{1}{2} \frac {\dot{\sigma^2_n}}{\sigma^2_n}
\mathcal{G} - \frac{1}{2}  \frac{\dot{\sigma^2_n}}{\sigma^2_n}
(\ad^n \!\!\cdot\! \zd - \ad^{0,n} \!\!\cdot\! \zd^{0})\mathcal{G}'
\nonumber \\&&+ (\dot\ad^n \!\!\cdot\! \zd
- \dot\ad^{0,n} \!\!\cdot\! \zd^{0})\mathcal{G}',
\label{appeq:gf-fp-2a}
\end{eqnarray}
where we have condensed some $z$ dependence by factoring out
$\mathcal{G}'$.  On the other hand, according to
Eqs.~\eqref{appeq:partial-G},\eqref{appeq:diff-G}, the r.h.s.\ of
\eqref{appeq:gf-fp-2} yields
\begin{eqnarray} \partial_t \mathcal{G} &=&  - \frac 12 (a_1^n)^2 \left( \frac
  {\mathcal{G}}{\sigma^2_n} + \frac{\ad^n \!\!\cdot\! \zd
- \ad^{0,n} \!\!\cdot\! \zd^{0}} {\sigma^2_n}\mathcal{G}' \right)\nonumber \\
&& - a_1^n  \mathcal{G}'  \sum_{k=1}^{n-1}
  \frac{a_1^k}{\sigma^2_k}
(\ad^k \!\!\cdot\! \zd - \ad^{0,k} \!\!\cdot\! \zd^{0}) \nonumber \\
&& - \sum_{k=1}^{n-1} z_k a^n_{k+1} \mathcal{G}'.
\label{appeq:gf-fp-2b}
\end{eqnarray}
Equating \eqref{appeq:gf-fp-2a} with \eqref{appeq:gf-fp-2b} should
give the sought after equations for $\ad,\ad^0,\sigma$. Comparing the
$z$-independent factors of $\mathcal G$ in \eqref{appeq:gf-fp-2a} and
in \eqref{appeq:gf-fp-2b} gives
\begin{eqnarray}
\dot{\sigma^2_n}=(a_1^n)^2.
\label{appeq:gf-fp-sigma}
\end{eqnarray}
Thus the full first lines on the r.h.s.\ of \eqref{appeq:gf-fp-2a} and
\eqref{appeq:gf-fp-2b} are equal.  In the rest we change the summation
variable $k$ to $l$ and then equate the respective factors of $z_k$ and
those of $z_k^{0}$ to obtain differential equations for the
coefficients
\begin{subequations}
\begin{eqnarray}
\dot a_k^n &=& - a_1^n \sum_{l=k}^{n-1} \frac{a_1^l
  a^l_k}{\sigma^2_l}  -a_{k+1}^n,\\
\dot a_k^{0,n} &=& - a_1^n \sum_{l=k}^{n-1} \frac{a_1^l
  a_k^{0,l}}{\sigma^2_l},
\end{eqnarray}
\label{appeq:gf-fp-a}
\end{subequations}
where we have used the condition that $a^l_k=a^{0,l}_k=0$ for $k>l$.
Next, we determine the time dependence of the $a$-s by assuming it to
be power law and requiring that terms in each differential equation
have the same power.  Thus we separate the time dependence, and for
later purposes also factorize the constants as
\begin{eqnarray}
a_k^n= t^{n-k} b^n_k \, c_n, \ \ \ a_k^{0,n}= t^{n-k} b^{0,n}_k\, c_n,
\label{appeq:gf-fp-bc}
\end{eqnarray}
for all nonnegative integers $n,k$. We also set $b_k^n=0$ for $n<k$ to
ensure that $a_k^n$ vanishes for such indices.  The $c_n$-s are made
unambiguous by requiring $b_1^n=1$ and, furthermore, since $a^n_n=1$
thus $b^n_n=1/c_n$.  Then \eqref{appeq:gf-fp-sigma} gives
\begin{eqnarray}
\sigma^2_n=(c_n)^2 \frac{t^{2n-1}}{2n-1}.
\label{appeq:gf-fp-sigma2}
\end{eqnarray}
The parameterization in \eqref{appeq:gf-fp-bc} is justified by the
fact that substituting it into \eqref{appeq:gf-fp-a} the $c$-s
disappear, so what remains are equations for the $b$-s as
\begin{subequations}
\begin{eqnarray}
b_{k+1}^n &=& -(n-k) b_k^n -  \sum_{l=k}^{n-1} (2l-1)b^l_k \\
b_k^{0,n} &=& - \frac {1}{n-k} \sum_{l=k}^{n-1} (2l-1)b_k^{0,l} .
\end{eqnarray}
\label{appeq:gf-fp-beta}
\end{subequations}
For a few small integer indices these equations can be solved, whence
the following general formulas can be surmised
\begin{subequations}
\begin{eqnarray}
b_{k}^n &=&  (-1)^{k-1} \frac {(n+k-2)!} {(n-k)!(k-1)!}, \\
b_k^{0,n} &=& (-1)^{n-k} b_{k}^n.
\end{eqnarray}\label{appeq:gf-fp-beta-solv}
\end{subequations}
Note that Eqs.~(\ref{appeq:gf-fp-beta}a) and (\ref{appeq:gf-fp-beta}b)
are homogeneous linear equations leaving room for overall factors in
the solution.  They are set by the conditions (i) $b^n_1=1$ and (ii)
$b^{0,n}_n=b^n_n$.  Condition (i) was stated earlier below
Eq.~\eqref{appeq:gf-fp-bc}, while (ii) is equivalent to the
requirement that $G_n$ depends on $z_n$ and $z_n^0$ only through their
difference.

One can confirm proposition \eqref{appeq:gf-fp-beta-solv} by
substituting it into \eqref{appeq:gf-fp-beta} and then using the
identities
\begin{subequations}
\begin{eqnarray}
\sum_{l=k}^{n-1} \!\frac {(2l-1)(k+l-2)!}{(l-k)!}  \hspace{-30pt} &&
\hspace{10pt}  = \!\! \frac{n-1}{k} \frac
{(n+k-2)!}{(n-k-1)!}, \\
\sum_{l=k}^{n-1} (-1)^l\frac {(2l-1)(k+l-2)!}{(l-k)!}
 \hspace{-60pt} &&\nonumber \\
 &=& (-1)^{n+1} \frac {(n+k-2)!}{(n-k-1)!},
\end{eqnarray}
\label{appeq:gf-fp-idents}
\end{subequations}
which may be proved by induction.

Finally, with \eqref{appeq:gf-fp-beta-solv} together with
\begin{eqnarray}
c_n &=& \frac {1}{b_n^n} =  \frac {1} {(-2)^{n-1} (2n-3)!!}
\label{appeq:gf-fp-c}
\end{eqnarray}
we have all ingredients of \eqref{appeq:gf-fp-bc} to calculate the
$a$-s, the result being displayed in Eq.~\eqref{eq:gf-parms}.  The
standard deviation $\sigma_n$ given in \eqref{eq:gf-parms} follows
then from \eqref{appeq:gf-fp-sigma2} and \eqref{appeq:gf-fp-c}.

\section{Leading perturbation of the PDF for large $\alpha$ from the
  $n$th mode} \label{app:large-alpha}

We start out from the Fourier representation~\eqref{eq:path-two-modes}
of the path with one mode of frequency $n$ beside the basic one
($n=1$). From the condition $h'(t_0)=0$, we obtain the correction in
the position $t_0=\pi/2+\delta_n$ of the maximum to leading order
\begin{equation}
  \delta_n=n\varepsilon_n\frac{a_n}{a_1}  \cos\left(\frac{n\pi}{2}+
    \varphi_n \right).
\end{equation}
Hence we can calculate the maximum to second order in $\varepsilon_n$
(the quadratic correction in $\delta_n$ contributes only to cubic
order)
\begin{eqnarray}
  h_\text{m}= h(t_0) &\approx& h(\pi/2) + h'(\pi/2)\delta_n + 1/2
  h''(\pi/2)\delta_n^2 \nonumber \\
  &\approx& a_1 + \varepsilon_n a_n \sin \left( \frac{n\pi}{2}+\varphi_n
  \right)\nonumber
  \\ && + \frac{n^2\varepsilon_n^2a_n^2}{2a_1}\cos^2\left(
    \frac{n\pi}{2}+
    \varphi_n\right).
\end{eqnarray}
Now the PDF for the MRH $h_\text{m}$ is obtained by averaging
over $a_1, a_n, \varphi_n$ in
\begin{eqnarray}
  P(z) &=& \left\langle \delta \left( z-h_\text{m}\right)
  \right\rangle \, .
\end{eqnarray}
Expanding to second order, one finds
\begin{eqnarray}
  P(z) &=& \left\langle \delta \left( z-a_1\right) \right\rangle -
   \left\langle \delta' \left(z-a_1\right)
    \left[ \varepsilon_na_n \sin \left( \frac{n\pi}{2}+\varphi_n
      \right)\right. \right.
  \nonumber \\
  && + \left. \left. \frac{n^2\varepsilon_n^2a_n^2}{2a_1}\cos^2\left(
        \frac{n\pi}{2}+ \varphi_n\right)\right] \right\rangle \nonumber \\
  && + \frac{\varepsilon_n^2}{2} \left\langle \delta''
    \left(z-a_1\right)
    a_n^2 \sin^2 \left( \frac{n\pi}{2}+\varphi_n \right)\right\rangle .
\end{eqnarray}
Note that here derivatives of the Dirac-delta appear. Now performing
the averages yields ($P_0$ is given by Eq.~\eqref{WD-pdf})
\begin{eqnarray}
  P(z) &=& P_0(z)+ \varepsilon_n^2 P_{2,n}(z),\label{appeq:pert-pdf-unsc1a}\\
  P_{2,n}(z) &=& - \frac{n^2}{2}\left( \theta(z) e^{-z^2}\right)' +
  \frac 12 \left( \theta(z)z e^{-z^2}\right)''.
  \label{appeq:pert-pdf-unsc1b}
\end{eqnarray}
Differentiation of the terms with step-functions gives
\begin{eqnarray}
  P_{2,n}(z) &=&  (1-n^2/2)\delta(z) +z \delta'(z)/2\nonumber \\
  && +e^{-z^2} \theta(z) \left( 2z^3 + (n^2-3) z\right),
  \label{appeq:pert-pdf-unsc2}
\end{eqnarray}
where the term proportional to $z^2\delta(z)$ has been omitted,
since it does not contribute to the average and other moments of
nonsingular functions.  Hence we obtain formula
\eqref{eq:pert-pdf-unsc-2}.

The mean to second order is best calculated from
\eqref{appeq:pert-pdf-unsc1a},\eqref{appeq:pert-pdf-unsc1b} resulting in
\begin{eqnarray}
  \left\langle z\right\rangle &=& \frac{\sqrt{\pi}}{2} +
  \varepsilon_n^2\frac{n^2\sqrt{\pi}}{4}.
  \label{appeq:pert-pdf-mean}
\end{eqnarray}
The scaled PDF is then obtained by the change of variable from $z$ to
$x=z/\left\langle z\right\rangle$, and expanding the resulting
expression to second order in $\varepsilon_n$
yields Eqs. \eqref{eq:pert-pdf-sc-1} and \eqref{eq:pert-pdf-sc-2}.

\bibliography{references}

\end{document}